\documentclass[sigplan,nonacm]{acmart}
\setcopyright{none}
\makeatletter
\let\@ACM@checkaffil\relax
\makeatother

\usepackage{algpseudocode}
\usepackage{algorithm}
\usepackage{subfig}
\DeclareCaptionFont{paperbody}{\normalfont\normalsize\rmfamily}
\usepackage{bytefield}
\usepackage{booktabs}
\title[Memory Fit]{Memory Fit: Architectural Support for Page
  Placement by Learning Models of Good Behavior}

\author{Chun Deng}
\affiliation{%
  \institution{Stanford University}
  \country{USA}
}
\email{dengchun@stanford.edu}

\author{Hui Sub Shim}
\affiliation{%
  \institution{Stanford University}
  \country{USA}
}
\email{hsshim@stanford.edu}

\author{Evan Laufer}
\affiliation{%
  \institution{Stanford University}
  \country{USA}
}
\email{emlaufer@stanford.edu}

\author{Anna Eaton}
\affiliation{%
  \institution{Stanford University}
  \country{USA}
}
\email{acbeaton@stanford.edu}

\author{David Tennenhouse}
\affiliation{%
  \institution{Independent Researcher}
}
\email{dtennenhouse@ieee.org}

\author{Tsachy Weissman}
\affiliation{%
  \institution{Stanford University}
  \country{USA}
}
\email{tsachy@stanford.edu}

\author{Philip Levis}
\affiliation{%
  \institution{Stanford University and Google}
  \country{USA}
}
\email{pal@cs.stanford.edu}

\renewcommand{\shortauthors}{Deng et al.}

\begin{document}

\begin{abstract}
  Server main memory is increasingly heterogeneous, incorporating
technologies such as zswap, CXL memory expanders, and emerging
phase-change memories (PCM). To decide where to place pages, operating systems
have to choose between efficient but coarse-grained access
bits, out-of-band profiling, or costly runtime instrumentation.

This paper proposes \emph{memory fit}, a hardware abstraction that provides
precise and accurate page placement information to an operating system.
Memory fit measures how well a page ``fits'' each type of memory in the system.
Memory fit separates the mechanism of measuring affinity between pages and memories from
the OS policy of choosing which pages to move when. The paper describes  
soft trace matching (STM), a computationally and space efficient algorithm to compute memory fit. 
The key technical insight
behind STM is that it can estimate memory fit by comparing accesses to a page
with a learned model of ``good'' traces for a memory.

We have implemented memory fit and STM in a BOOMv3 RISC-V processor
running Linux 6.6.
On extended GUPS, memory fit placements recover 2.2--3.8$\times$ as much of
the performance gap to optimal placement as Linux TPP. For Redis with equal
DRAM and PCM capacities, placements selected by memory fit recover 80\% of the
performance gap, versus 19\% for TPP, in a placement replay. Memory fit requires only 64 bits of state per
4\,KiB page and can be updated on every main-memory access with only a 1.2\%
overhead of OS kernel notification processing.

\end{abstract}

\maketitle

\section{Introduction}
\label{sec:introduction}

DRAM has grown to dominate the cost of modern servers~\cite{tpp,tmts,microsoftmem}.
Reducing this cost has motivated systems that combine several forms of memory.
These include software-compressed memory such as zswap, CXL memory expanders,
emerging phase-change memories and High Bandwidth Flash~\cite{zswap,ocp-cxl,zilberberg13pcm}.
Each provides higher capacity per dollar (GB/\$), with different latency, bandwidth, density, and
read/write behavior.

The performance differences of these memories create a placement problem for an
operating system: which pages should it place in each memory? Existing systems
generally answer this question by measuring how frequently a page is accessed,
describing it as page temperature. Hot pages are accessed frequently while cold
ones are accessed rarely. Transparent Page Placement (TPP), for example,
proactively moves cold pages out of local memory and reactively promotes hot
pages~\cite{tpp}.  HeMem and Memtis improve how the operating system measures
and classifies page temperature~\cite{hemem,memtis}.

Temperature-based policies work well when there is a hierarchy of
slow to fast memories: hotter pages benefit more from faster
memory. As memory performance has become more complex, however, temperature
is too coarse: two pages with the same temperature can observe
different performance in the same memory. For example, 
memory-level parallelism means sometimes hot pages are better put in slower CXL
memory~\cite{beyondhotness}, compression within a
memory expander can make access cost depend on the contents of a
page~\cite{park26cxl}, and PCM can respond differently to reads, writes, and their
ordering~\cite{zilberberg13pcm}.

Processors today provide poor support for an operating system to make page
placement decisions. Page-table entries provide accessed and dirty
bits\footnote{Some ARM processors lack hardware dirty-bit
management~\cite{arm-dirty-bit}.}, which are efficient but coarse-grained,
while more detailed mechanisms such as Intel PEBS and AMD IBS require sampling
and impose a large software processing cost.\footnote{NEMO~\cite{nemo} explains
the tradeoffs of memory measurement excellently.} Hardware--software codesign
approaches such as NEMO~\cite{nemo} and M5~\cite{m5} show how better hardware
support for performance counters can substantially improve placement.

This paper takes a step back to ask two broader questions. How should a
processor support an operating system's page placement policy? What
abstraction should it provide? Examining the tradeoffs of different memory
technologies and how page placement affects system performance,
Section~\ref{sec:requirements} derives three requirements for such an
abstraction: it must be expressive enough to distinguish many different
memories and access patterns, it must provide a very simple interface to
software, and it must be fast enough to precisely measure accesses without
delaying memory requests.

Section~\ref{sec:design} describes memory fit, a hardware abstraction that
meets all three requirements. A memory fit score measures how
well a page's accesses match patterns that perform well in each memory. A
64-bit per-page record accommodates up to four eight-bit memory fit scores
with a compressed access history.
Memory fit scores are comparative. Comparing one page across memories indicates
which memory is best for that page, while comparing page scores for a memory
indicates how to allocate limited capacity. Memory fit separates
mechanism from policy: hardware calculates an affinity signal between a page's
accesses and each memory, while the operating system implements a policy that
combines this signal with capacity, migration cost, and application constraints.

The primary research challenge that arises is computing memory fit precisely
and efficiently with very limited state.
Section~\ref{sec:stm-computation} proposes soft trace matching (STM), a
novel algorithm for computing memory fit. STM represents accesses as a
sequence of reads and writes with normalized reuse distances. Building on
decades of work in extremely fast hardware compression, its simplest form is
a single table of learned weights; additional tables condition 
on history. The key takeaway
is that, drawing from the information theory of compression, STM
can learn extremely concise and accurate models for memory fit by 
seeing what  ``good'' traces for each memory look like.
STM measures the affinity between a trace and memory as the relative
completion-time benefit of placing the same work in each memory.

We have implemented memory fit and STM in a BOOMv3 RISC-V FPGA processor
and incorporated support in Linux 6.6. Using this implementation, we
evaluate STM against Linux TPP using extended GUPS, Redis, and a memory-access
workload derived from llama.cpp.
On extended GUPS, STM recovers 72.4\% of the optimal-to-worst placement cycle
gap under static behavior and 71.9\% when behavior reverses midway, compared
with 19.2\% and 33.3\% for TPP.  Across
three Redis capacity ratios, STM recovers 37.1--90.4\% of the
all-slow-to-all-fast performance gap, while TPP recovers 7.9--31.8\%.

This paper makes four contributions:
\begin{itemize}
  \item It defines memory fit as a comparative signal for placing pages across
  memories with different performance characteristics.

  \item It presents STM, a learning-based algorithm for computing memory fit from page
  access patterns.

  \item It presents and implements a hardware--software architecture that
  exposes memory fit to Linux while leaving placement policy in the operating
  system.

  \item It evaluates STM against Linux TPP and in combination with existing
  placement policies.
\end{itemize}

Section~\ref{sec:motivation} describes memory fit.
Sections~\ref{sec:architecture} and~\ref{sec:implementation} describe its
architecture and implementation.  Section~\ref{sec:eval} evaluates a complete
Linux system with memory fit,
Section~\ref{sec:related} discusses related work, and
Section~\ref{sec:conclusion} concludes.

\section{Motivation}
\label{sec:motivation}

A page-placement mechanism must solve two problems:
provide a useful
signal on how a page would perform in each memory, and
compute that signal efficiently for many pages.
Existing systems have made substantial progress on efficiency, but
do so by providing a low-quality signal based on frequency/recency or calculating
offline signals that assume program invocations do not change significantly.

\subsection{Memory Use Varies Greatly in Space and Time}

\begin{figure}[t]
  \centering
  \includegraphics[width=\columnwidth]{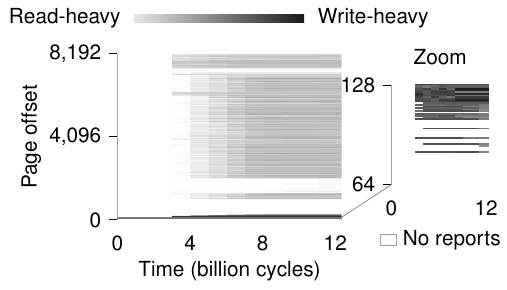}
  \caption{STM scores vary across Redis pages and over time.}
  \Description{A full 8,192-frame grayscale heatmap with thin individual-page
    rows, alongside a zoom of offsets 64 through 127. Both show the same
    12.35-billion-cycle interval and grayscale scale. Several zoomed rows
    become lighter later in the run. White means no valid report.}
  \label{fig:memory-fit-heatmap}
\end{figure}

Figure~\ref{fig:memory-fit-heatmap} shows a diagnostic capture of how Redis, a
key-value store, accesses physical pages over time; it was gathered using our
custom RISC-V FPGA processor described in Section~\ref{sec:implementation}.
Because of the virtual-to-physical mapping, adjacent physical frames can have
different access patterns: some are read-heavy, others are write-heavy, and
some are accessed infrequently.  Furthermore, individual pages change in the
timescale of hundreds of millions of cycles.  

These changes over space and time
motivate the need for software to have accurate per-page information that
quickly updates with access pattern shifts. The exact policy of when to move
pages and which pages to move can remain in software, but page measurement
mechanisms require hardware support.

\subsection{Hotness Does Not Identify the Best Memory}

Page temperature (access frequency) is a good signal when one memory is strictly
better than another.  If every access to a better memory has roughly the same
benefit, hotter pages should be placed in the better memory.  Prior research
and software systems differ in how they measure page temperature, but they
ultimately rank pages by access activity.  Transparent Page Placement
(TPP)~\cite{tpp} uses recency, HeMem ranks pages by sampled access frequency
and uses write counts only to break ties~\cite{hemem} and Memtis adapts the
hot-set threshold to the observed access distribution~\cite{memtis}.  M5, a
hardware-software approach, moves ranking into the CXL controller with
fixed-size trackers for the most frequently accessed pages and words~\cite{m5}.

Newer memory technologies break the assumption that accesses have mostly
uniform costs. Hot pages with parallel accesses are better suited to higher
latency CXL memory than colder pages with serialized
accesses~\cite{beyondhotness}.  CXL memory expander compression means two pages
exhibit different latency~\cite{park26cxl}.  Phase change memory (PCM) write
costs depend on the preceding operation~\cite{zilberberg13pcm}.

\subsection{Observing Memory Is Expensive}

Estimating where a page should be placed requires measuring how that
page is accessed.  Page-table access and
dirty bits provide broad coverage, but recovering recent information
requires scanning page tables, clearing state, and synchronizing
address translations.  The work grows with the amount of memory
examined rather than with the number of pages that need a placement
decision. 

The tension is that more precise and higher frequency observations
allow an OS policy to make better decisions. Research systems
such as HeMem and Memtis use low-level memory event counters
from Intel PEBS to gather richer information than access
and dirty bits. 

Detailed measurements are costly at modern memory capacities.  
Memory events from PEBS or AMD IBS require
probabilistic downsampling to curtail the CPU overhead of processing
them, reducing the precision and accuracy of these
estimators.~\cite{nemo}
HeMem reports
that page-table scanning reduces GUPS throughput by 18\% relative to
its PEBS configuration, and that aggressive PEBS settings drop up to
30\% of samples when the processing thread falls behind~\cite{hemem}.
Memtis adjusts its PEBS period dynamically to keep sampling
below 3\% of one core~\cite{memtis}.  NEMO summarizes the same
tradeoff more broadly: software processing forces trading off
coverage, timeliness, and granularity for overhead, and prior work
reports more than 50\% application slowdown when sampling 10\% of
cache misses~\cite{nemo}.

NEMO finds
that its more precise memory-controller counters find more than twice
as many skewed huge pages, identify candidates up to 10.4$\times$
faster, and improve throughput by 13\% at the most constrained
fast-to-slow capacity ratio~\cite{nemo}.
Unfortunately, maintaining per-page counters in on-chip SRAM, as NEMO does, is
prohibitively costly.
Its Memtis configuration uses 1.4\,MiB of SRAM and takes approximately
37 seconds to sweep 16\,GiB at 4\,KiB granularity.  At the same
SRAM-to-memory ratio, a 1\,TiB memory requires approximately 90\,MiB
of SRAM~\cite{nemo}, which is larger than the per-die caches in modern
multi-chiplet processors.

\subsection{Requirements for Architectural Support}
\label{sec:requirements}

The complexities described in this section motivate three requirements for  
architectural support of operating system page placement policies:

\begin{enumerate}
\item \textbf{Expressive.}  Hardware should provide a mechanism that
  distinguishes the performance tradeoffs of different memories while
  supporting new memory technologies.

\item \textbf{Simple.}  The mechanism should expose simple per-page
  values that OS policies can easily use without knowing the details of
  the different memories in the system.

  \item \textbf{Efficient.}  The mechanism must maintain these values with bounded
    on-chip work, without delaying demand memory requests, and without
    requiring large amounts of on-chip memory.
\end{enumerate}

\begin{table}[t]
  \centering
  \small
  \setlength{\tabcolsep}{1.5pt}
  \newcommand{\tblcheck}{\(\checkmark\)}
  \newcommand{\tblchecklimited}{\(\checkmark^{*}\)}
  {\small
  \begin{tabular}{@{}llccc@{}}
    \toprule
    System & Placement signal
      & Expressive
      & Simple
      & Efficient \\
    \midrule
    TPP~\cite{tpp}       & Temp (hint faults) &         &           & \tblcheck \\
    HeMem~\cite{hemem}   & Temp (PEBS)     &           &           &           \\
    Memtis~\cite{memtis} & Temp (PEBS)     &           &           & \tblcheck \\
    M5~\cite{m5}         & Temp (counter)  &           & \tblcheck & \tblcheck \\
    NEMO~\cite{nemo}     & Temp (counter)  & \tblcheck & \tblcheck &           \\
    \textbf{Memory Fit}  & STM             & \tblcheck & \tblcheck & \tblcheck \\
    \bottomrule
  \end{tabular}
  }

  \caption{Properties of page placement systems.}
  \label{tab:placement-comparison}
\end{table}

Table~\ref{tab:placement-comparison} shows how prior systems meet these
criteria. Software solutions like TPP, HeMem, and Memtis require new
algorithms/software support for each memory technology, and push this
complexity into the OS. Hardware solutions such as M5 and NEMO
provide a simple interface to software, but M5 is limited to access counts
and NEMO has substantial on-chip state and limited programmability.
The next section describes memory
fit, which meets all three requirements.

\section{Architecture}
\label{sec:architecture}
\label{sec:design}

Memory fit is a hardware abstraction to assist page
placement decisions. It expresses how well
a page's observed access pattern suits each available memory. Hardware computes
memory fit from the post-cache access stream using Soft Trace Matching (STM),
a novel algorithm that draws on decades of work in fast hardware
compressors to match accesses against good access patterns for a memory
(Section~\ref{sec:stm-computation}). STM learns offline a concise and efficient
model of what good traces for each memory look like.
Hardware exposes scores through a memory-backed PFN table and notifies the
operating system through a ring of physical frame numbers. The OS decides
how and when to act on these scores under capacity, migration, and
application-policy constraints. 

\begin{figure}[t]
  \centering
  \includegraphics[width=2.8in]{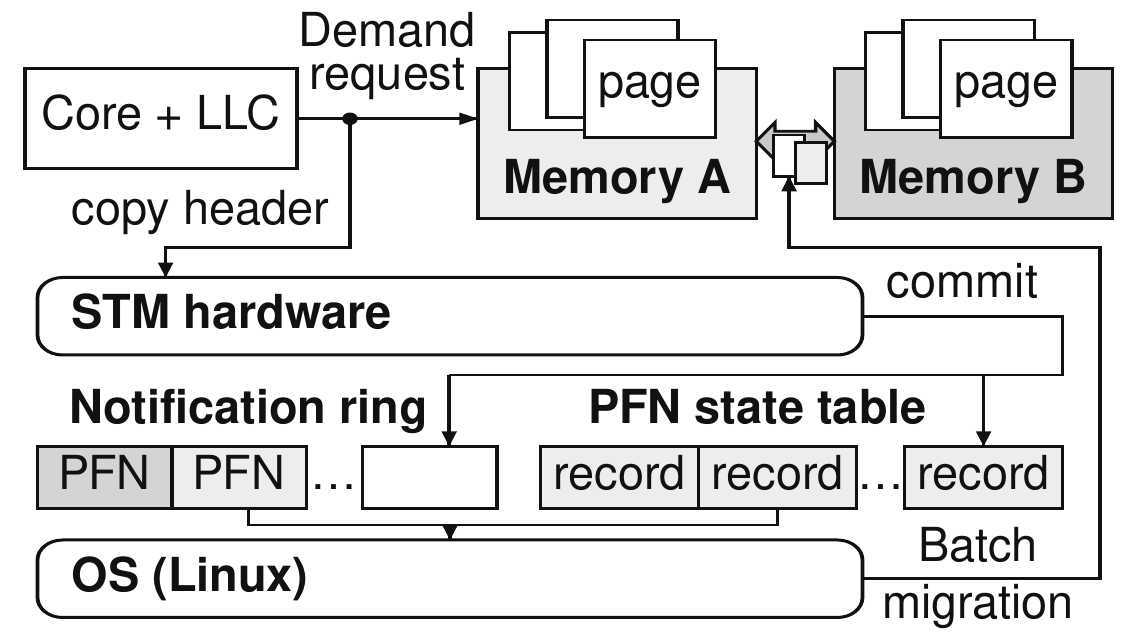}
  \caption{Memory Fit architecture. STM publishes records and notifications
  in coherent memory; Linux migrates pages to their preferred memories.}
  \Description{A demand memory request travels from the core and last-level cache to
  memory while a side branch copies its header to STM hardware.  STM hardware
  writes a notification ring and PFN state table.  The OS (Linux) reads these shared arrays
  and moves pages between Memory A and Memory B.  The internal hardware and
  software components are shown in separate detail figures.}
  \label{fig:stm-architecture}
\end{figure}

\subsection{What Is Memory Fit?}
\label{sec:fit-semantics}

Memory fit indicates which memory is likely to better serve a page's
observed access sequence. In this paper, better means faster. Other
training objectives could target metrics such as energy; we defer
a discussion of this generality to Section~\ref{sec:conclusion}.

A low fit score means
that the page's recent access sequence should perform well on that memory.
Scores for different memories are comparable: for one page, a lower score
indicates a better memory, and the difference expresses the benefit of one
placement over the other. Differences are also comparable across pages:
a lower score indicates a better page for a memory. The fact that fit
scores are comparable means an OS can prioritize pages with the largest reduction
in memory fit score. Readers interested in a formal specification of memory
fit, its objective, and probabilistic interpretation can refer to
Appendix~\ref{app:formal}.

\subsection{Example Operation}

Figure~\ref{fig:stm-architecture} follows one request through the memory fit
architecture.  A monitor sits after the last-level cache, observing requests to
main memory.  When the monitor accepts a request, it generates an event tuple,
evaluates each memory's model, and sends the resulting fit score updates to a
small on-chip delta cache.  The delta cache accumulates updates for each page
until it writes the page's fit metadata back to a main memory table indexed by
physical frame number.  When hardware commits a record that requires software
attention (e.g., the preferred memory changed), it pushes the physical frame
number into an event ring.  The operating system pulls frame numbers from the
ring and reads the corresponding fit metadata to update its policy state on
which pages to move and when.  The remainder of this section describes each
stage of this processing in greater detail.

\subsection{Soft Trace Matching}
\label{sec:stm-computation}

Soft trace matching (STM) computes memory fit as a weighted combination
of how well sequences of observations match those in its learned model. 
It takes a stream of observations that record
whether an access is a read or a write and its normalized reuse distance.
The normalized reuse distance is the number of other accesses since the page's
previous observation, divided by the current working-set size. As working-set
sizes can vary by several orders of magnitude, normalization provides two
benefits. First, it expresses these large differences in absolute reuse
concisely. Second, it allows a single model to work across different
working-set sizes.

STM scores an access with each memory's learned weights. Hardware keeps a
running tally for each page and memory, averaging these scores with more weight
on recent accesses.  The first table uses the current observation alone,
distinguishing reads from writes and short reuse from long reuse.  Additional
tables make this a \emph{conditional tally} by considering sequences of two or more
observations, distinguishing a write after a read from a write after a write. Such history
matters when, for example, a PCM write first reads the media before updating
it~\cite{zilberberg13pcm}. Our implementation of STM, described in
Section~\ref{sec:implementation}, uses two tables, so estimates condition on
only a single prior observation. We have found two tables sufficient in
practice. For simplicity, the rest of this section assumes there are only two
tables, but the statistics and hardware can be easily generalized to more.
Section~\ref{sec:training} describes how these tables are learned offline; they
remain fixed during execution and are shared across pages.

\begin{figure}[t]
  \centering
  \includegraphics[width=3in]{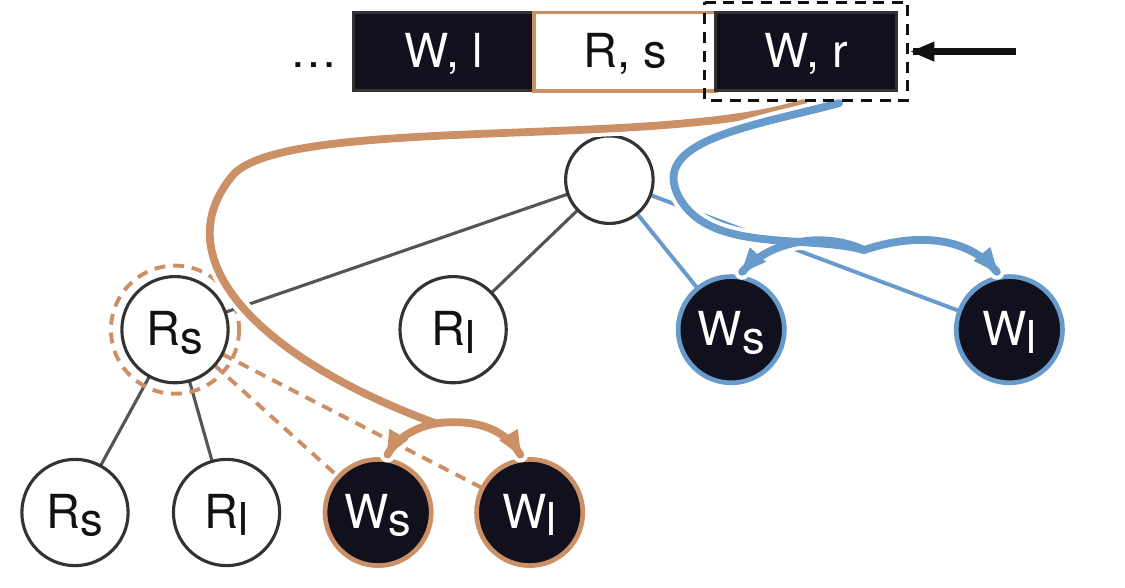}
  \caption{Soft matching on STM's conditional counters.
Hollow
and filled circles denote reads and writes; $s$ and $l$ identify learned short-
	and long-reuse anchors. The resulting score is a weighted sum of the
	interpolation between orange values and between blue values.}
  \Description{A page's access stream contains a short-reuse read followed by
  a write above a tree of circular read and write entries. The root's four
  children represent marginal counters. Only the preceding short-reuse read
  context expands into another four children. A blue curved arrow carries the
  current write to the root's two write anchors; an orange curved arrow carries
  it to the conditional pair. Both pairs are interpolated for the fit update.}
  \label{fig:stm-lookup}
\end{figure}

STM's tables draw on the nested organization of context tree weighting
(CTW)~\cite{ctw}, a lossless compression model representation. STM extends CTW
by interpolating between reuse anchors and weighted combination between
tables because memory access traces require probabilistic matching: otherwise
tables would require every possible combination of reuse distances. Observations below
the low reuse distance and above the high reuse distance are directly matched
to the anchors. Using the
same nested context organization allows STM to draw on the techniques
used for extremely fast CTW hardware compressors.
Section~\ref{sec:reuse-estimation} describes one such implementation.
Appendix~\ref{app:trace-model} gives the complete tree in
Figure~\ref{fig:stm-inference} and the corresponding equations.

Figure~\ref{fig:stm-lookup} shows an example of STM processing 
a write after a short-reuse read. 
We call the reuse distances in model table entries
``reuse anchors'' to distinguish them from the reuse distances of incoming memory
requests. The root's children count the current event alone;
the expanded $R_s$ subtree conditions those counts on the preceding short-reuse
read. The blue curved arrow matches the write to the root's two $W$ entries;
the orange arrow matches it to the conditional pair. Hardware blends each
pair's learned weights according to the write's reuse distance $r$, then
combines the two results to score the access.

\subsection{Updating Memory Fit in Hardware}
\label{sec:hardware-fit}
\label{sec:bounded-observation}

\begin{figure}[t]
  \centering
  \includegraphics[width=2.8in]{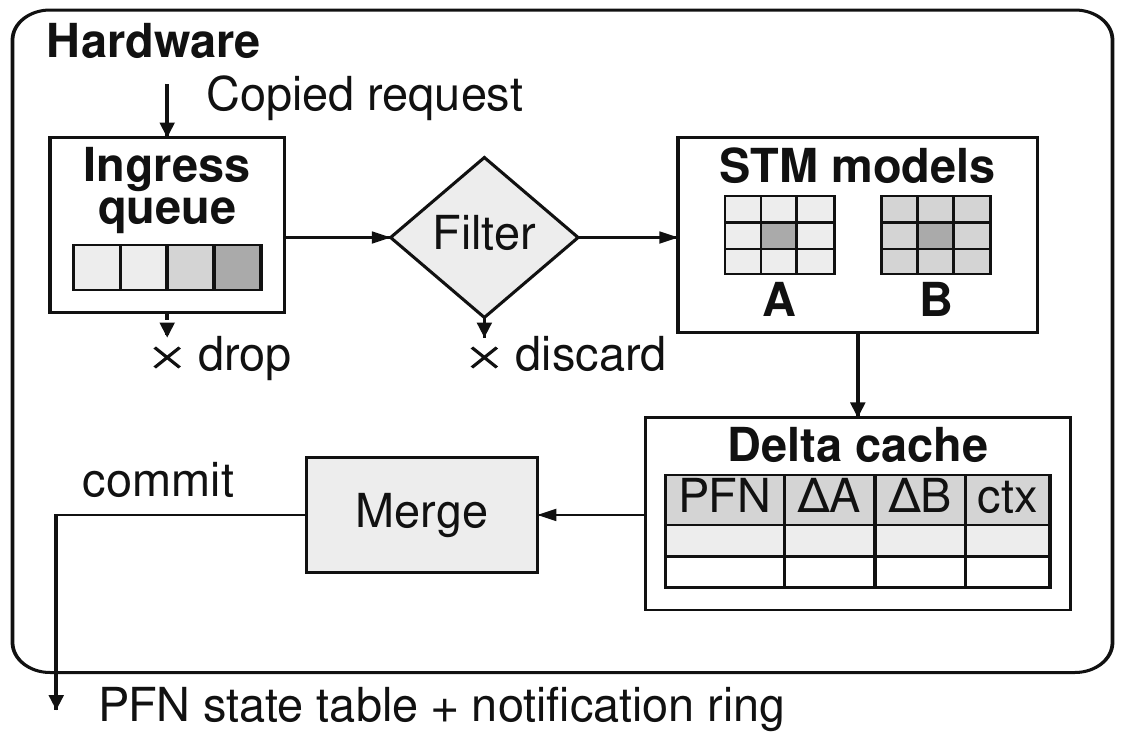}
  \caption{STM hardware accumulates observations in the delta cache and
  commits scores and notifications through coherent memory.}
  \Description{Copied requests enter an ingress queue, pass through a filter,
  and evaluate one STM model per memory.  Updates accumulate in the delta cache.
  The merge engine commits to the PFN state table and publishes notifications
  in the event ring. The flow continues from the first row
  to the second row.}
  \label{fig:stm-hardware}
\end{figure}

Hardware applies the STM computation in Figure~\ref{fig:stm-lookup} on
each post-cache request (Figure~\ref{fig:stm-hardware}). For each
observation, it computes one memory fit update per memory and
accumulates the updates in the on-chip delta cache. On write-back, one
non-atomic read-modify-write operation merges these deltas with the
memory-backed PFN record.  The design maintains three important architectural properties:

\noindent\textbf{Non-blocking and off-critical-path.}
The monitor copies request headers into an ingress queue while memory serves
the demand memory requests. If the queue is full, hardware drops the copied header
and continues serving the request. STM computation therefore runs off the
demand path and generates no back pressure. The queue can reduce observation
coverage under high pressure. Section~\ref{sec:observation-pipeline} describes
how the implementation bounds the resulting metadata work.

\noindent\textbf{Bounded metadata traffic.}
The delta cache has a fixed number of entries and combines repeated updates to
the same page into a single non-atomic read-modify-write operation.  Before a request
enters the STM pipeline, the range filter removes requests to the PFN table and
event ring so memory requests generated by STM do not recursively
generate more STM requests.  When the write-back path begins to fill, the
delta cache reduces the rate at which it admits pages that are not already
resident.  This feedback reduces the metadata request rate until the
write-back path recovers, while requests for resident pages continue to
accumulate in the delta cache.

\noindent\textbf{Eventual consistency.}
The write-back engine commits the complete PFN record before it appends the PFN
to the event ring. 
The commit engine serializes PFN updates.
Therefore, a shared monitor for multiple cores or several replicated monitors
can feed the same commit engine without creating separate versions of a page's
memory fit state.
The ring entry only identifies the page and does not
contain a copy of its memory fit record. Thus, the OS always reads the latest
committed record and revalidates the page before acting on the notification.
If a notification is delayed or coalesced with a later notification, the OS may
perform a useful migration later, but it does not act on 
stale information.

\subsection{Communicating Memory Fit to Software}
\label{sec:shared-state}

At boot, the operating system uses MMIO to specify the ranges of physical memory
to track, where to place the memory fit table and ring buffer, and enables
the memory fit hardware.  Once
STM is enabled, the write-back engine writes per-page memory fit state to the
coherent PFN table and appends page identifiers to the coherent event ring.
Both the memory fit records and the notifications therefore move through
ordinary coherent memory. Therefore, although the prototype uses one core, we designed this
interface to address coherent sharing across cores. MMIO is not used to transfer per-request or
per-page state during execution.

Each PFN has a 64-bit record (Figure~\ref{fig:record-layout}).
The format can accommodate up to four eight-bit memory fit scores. Our prototype,
described in Section~\ref{sec:implementation}, uses two memories and two score
bytes. The timestamp, count, and
context fields allow the STM computation to resume after the delta-cache entry
has been written back and later reinstalled.  The remaining fields store the
notification state and an optional coarse memory-level-parallelism (MLP) class
for the page.  At eight bytes for each 4\,KiB page, the PFN table occupies
0.195\% of the memory it covers.

When hardware publishes a PFN, it marks the record as having an outstanding
notification.  Further updates to that record do not append duplicate ring
entries until the OS consumes the current entry.  The ring raises an interrupt
after it collects a programmable batch or after a timeout expires.  This
allows Linux to process notifications in batches under load without leaving
an isolated notification in the ring indefinitely.

The interrupt interface, combined with the update ring, is
scanless: software does not need to scan the memory fit record
table. It only considers pages placed into the ring.

\section{Implementation}
\label{sec:implementation}

We have implemented memory fit and STM as a FireSim~\cite{firesim} target with a Small
BOOMv3~\cite{boomcore} core, a one-wide out-of-order RISC-V processor. The core
runs on an AWS F2 FPGA and boots Linux 6.6. This section describes important
and non-obvious details in the hardware implementation as well as how Linux
interacts with the memory fit architecture described in
Section~\ref{sec:architecture}.

\subsection{Core Configuration and Parameters}

The core's L1 instruction and data
caches are each 16\,KiB and four-way set associative.  The LLC is 4\,MiB,
eight-way associative. The core has two physical memories with 16\,GiB of
total addressable capacity.  Both memories use a quad-rank, FR-FCFS timing model, and
Linux sees them as two contiguous 8\,GiB NUMA nodes.
The target core and memory bus operate at 1\,GHz, while the remaining target
buses and STM operate at 500\,MHz.  Because FireSim executes the target at a
lower FPGA-host frequency, the routed FPGA image runs at 18\,MHz.

\subsection{Observation and Delta-Commit Pipeline}
\label{sec:observation-pipeline}

At the LLC output, a TileLink bus adapter adds each memory-request header into an 
ingress queue.  If the queue is full, the adapter drops the header and continues
serving the demand request.  Address filters discard
requests outside tracked memory ranges
as well as accesses to the state
table and event ring.  Excluding these metadata accesses prevents STM from
recursively observing its own memory accesses.  A separate TileLink bus client issues
coherent accesses to the PFN table and event ring without involving the core
pipeline.

The STM implementation
uses a 64-entry, four-way
delta cache.  A hit updates the delta-cache entry.  A miss evicts an
entry if needed and creates a new entry at zero, so it does not have to read a record
from main memory.
Starting an entry at zero is possible because updates are a decaying average so
are associative: an update can be merged with a record on eviction. The
delta cache entry has two terms: how much of the prior memory fit score remains after exponential
decay, and the evidence accumulated in the entry.

At commit,
hardware reads the memory-backed record once, applies both terms, and writes the
result.  This read-modify-write is non-atomic.  It produces the same decaying
average as applying each observation in order, even though hardware did not
read the PFN table on the initial cache
miss.  The acknowledged write is therefore the commit point.  The merge engine
serializes commits to the same PFN before it publishes a notification.

Because evictions access memory-backed metadata, the commit queue
also controls how much traffic STM can generate.  If coherent writes fall
behind, its occupancy lowers the probability that a new page evicts an
existing entry.  Pages already in the cache continue to
accumulate updates, but evictions and associated table accesses fall
as pressure rises.

\subsection{STM Computation and Approximations}
\label{sec:reuse-estimation}

The STM datapath begins by combining operation type (read or write) with a
normalized reuse distance measured relative to the current working set.  An
eight-bit timestamp advances after a fixed number of observed requests.  For a
page that remains in the delta cache, subtracting its previous timestamp
estimates how many memory-side accesses occurred between references.

One wrinkle is that the first observation after a delta-cache installation has no
page-local timestamp because installation deliberately avoids reading the PFN
table.  Hardware uses a global reuse estimate for this observation.  At commit,
hardware compares the timestamp in the old PFN record with the first timestamp
stored in the new delta entry.  These samples update the global estimate used
by later installations.

In parallel, a 16-register HyperLogLog estimator~\cite{hyperloglog} estimates
the number of distinct pages in the current working set.  Dividing reuse
distance by this estimate places reuse on a working-set-relative scale and allows the same
learned anchors to apply across different working-set sizes.
Section~\ref{sec:stm-computation} describes the interpolation and table lookups.

Two randomized mechanisms preserve information with limited state.
First, the affine transform retains fractional precision, but
the memory-backed record holds only eight bits per memory fit score.  Truncating a
sub-unit change would prevent repeated small updates from
accumulating.  Stochastic rounding instead chooses between the two adjacent
integer values according to the fractional part.  Both memories share
the same random draw so rounding cannot introduce an artificial ordering
reversal.

Second, randomized tickets are a low-pass filter when two memory fit
scores are close. Small updates can repeatedly flip ordering,
making the OS track noise; a low pass filter removes this higher
frequency signal.
Appendix~\ref{app:quantized-updates} explains how compact tickets and the
page's stability level filter these reversals at commit.

An optional memory-level-parallelism (MLP) extension counts outstanding
memory reads and maps the count into four classes.  Linux may use the class
to refine the order of candidates with similar memory fit benefits.  The MLP
class remains a separate ranking input and does not alter STM computation or
either memory fit score.

\subsection{Training and Model Quantization}
\label{sec:training}

The per-memory models STM uses are learned offline. Our training pipeline
takes a large set of synthetic fixed-work access traces and simulates their completion time on each
memory. This identifies which memory is faster for each trace and by how
much.

The goal of STM models is to quantitatively distinguish different memories and
the tradeoffs for arbitrary access patterns. Therefore the training set
consists of 200 traces that have different read/write mixtures, reuse scales,
conditional contexts, locality, concurrency, warm and cold state, and phase
changes.  These traces have no connection to the applications we evaluate
in Section~\ref{sec:eval}: they were chosen before we selected
these workloads.

We implement STM as a differentiable function so the training framework can
adjust the probability tables, reuse anchors, short/long-reuse boundary, and
mixture weights together. Training makes the faster memory receive a lower
memory fit score and makes the score gap reflect its relative completion-time
advantage. It also preserves the ordering of access patterns by this advantage.
Pages that gain little from fast memory have similar scores for both memories;
pages that gain more have larger gaps. The OS can therefore reserve scarce
fast-memory capacity for the pages that benefit most. Access frequency
contributes through reuse and its effect on the simulated completion time.

Hardware stores table entries and weights as eight-bit values. Training
accounts for this limited precision, then rounds the learned constants to the
hardware representation and checks the resulting integer model. The prototype
compiles these constants into its hardware image. Programmable tables could
instead load another trained model at boot.
Appendix~\ref{app:training-details} specifies the objective and quantization.

\subsection{Coherent Interface and Linux Driver}
\label{sec:software-interface}

\begin{figure}[t]
  \centering
	\includegraphics[width=\columnwidth]{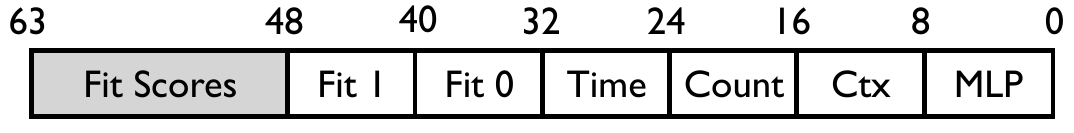}
  \normalsize
  \caption{64-bit PFN record. Ctx is the STM context and
	MLP stores the bits for memory parallelism and record stability.
	The prototype has 2 memory fit scores but the reserved top 16 bits
	can hold an additional two.}
  \Description{A row divides a 64-bit record into reserved space, two
  memory fit scores, a timestamp, an observation count, context and control
  state, and MLP and stability state.}
  \label{fig:record-layout}
\end{figure}

At boot, the driver allocates a coherently shared state table and a
4,096-entry event ring.  It sets their physical addresses, the two tracked
address ranges, and policy parameters through MMIO.  Observations begin only after both
shared structures are initialized.

Hardware completes writing a PFN record before publishing the corresponding PFN
to the event ring.  An outstanding-notification bit prevents a page from
occupying more than one unread ring entry.  Hardware atomically commits the
new record while setting this bit.  When Linux drains that entry, it atomically
obtains the current record while clearing the bit.  This protocol coalesces
repeated changes without losing the final state.

The two tracked ranges use dedicated movable pageblocks so managed pages
remain eligible for Linux's migration path.
Before migration, the driver commits any resident page delta.  After a
successful copy, it transfers the 64-bit record to the destination PFN and
clears the source.  A failed copy retains the source record.  The driver
serializes this transfer with hardware commits so the page has one
authoritative record before and after migration.

Appendix~\ref{app:implementation-details} gives the full implementation,
including further driver and record details.

\subsection{Acting on Memory Fit in Software}
\label{sec:placement-loop}
\label{sec:fitd}

\begin{figure}[t]
  \centering
  \includegraphics[width=2.8in]{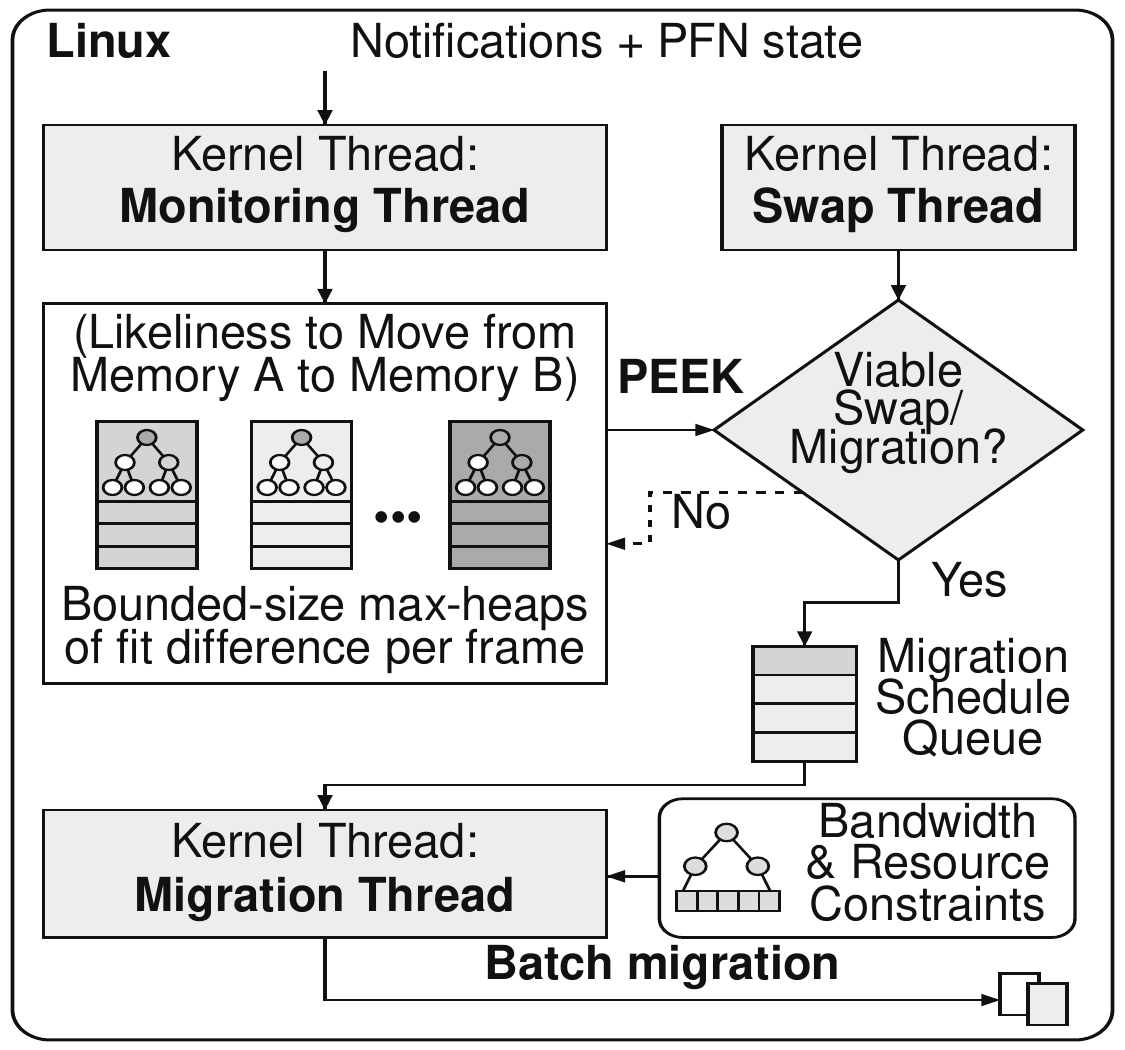}
  \caption{Linux prioritizes page moves in bounded heaps and schedules
  migrations under bandwidth and resource constraints.}
  \Description{The monitoring thread updates bounded max-heaps.  The swap
  thread checks feasible moves and feeds a migration schedule queue.
  Rejected moves return to the heaps.  A migration thread executes batches
  under bandwidth and resource constraints.}
  \label{fig:stm-linux}
\end{figure}

Figure~\ref{fig:stm-linux} shows the software placement path.
The driver divides interrupt processing into an upper half and a lower half.
The upper half checks the interrupt source, disables further ring interrupts,
and schedules the lower half before returning.  The lower half runs in a
high-priority workqueue, drains the available notifications, and computes the
benefit of moving each eligible page.  It then acknowledges the interrupt and
reenables notification delivery.

The lower half maintains one bounded max heap for pages currently in each
memory.  A heap key is the reduction in total memory fit score obtained by
moving that page to another memory.  A later notification updates the existing
entry for the page instead of adding a duplicate.  When a heap is full, a new
page replaces a weaker candidate only when it offers a larger improvement.
Thus, the heaps retain a bounded working set of the best moves reported by
hardware.

A separate migration thread wakes after the lower half changes a heap.  If the
destination has free capacity, the thread selects the page with the largest
benefit.  If the destination is full, it pairs the best incoming page with a
resident page from the opposite heap and accepts the swap only when the pair
reduces the combined memory fit score.  The thread revalidates each selected
page, groups accepted pages into bounded batches, and invokes Linux's existing
page-migration path.

The software placement path maintains three properties to support scaling to large numbers of pages:
\begin{enumerate}
  \item \textbf{Scanless.} The driver processes only notified pages and never
  scans memory or recomputes a global placement.
  \item \textbf{Incremental.} Each notification requires $O(\log k)$ work in a
  bounded $k$-entry heap, and each migration cycle processes a bounded batch.
  \item \textbf{Convergent.} For fixed memory fit scores, each accepted move or
  swap improves the placement until no represented candidate can improve it
  further.
\end{enumerate}
Appendix~\ref{app:matching} formalizes the local-minimum property and the
conditions under which greedy matching is at least one half of the global
optimum.

\section{Evaluation}
\label{sec:eval}

This section evaluates how memory fit aids OS page placement, its overheads,
and the implications of some of the implementation design decisions. It
does so by answering four questions:

\begin{itemize}
 \item Does memory fit accurately guide page
	 placement?

 \item What overheads do memory fit observations and page
	 migrations introduce, and do they impact the
		performance benefits of better placement?

 \item Does memory fit
improve application performance across different capacities for memory types
		and different memory setups?

 \item Does memory fit compose with complementary placement
	 techniques such as Colloid and per-page MLP?

\end{itemize}

\subsection{Methodology}

We use the FireSim prototype described in Section~\ref{sec:implementation} and
compare STM with Linux TPP~\cite{tpp}. The goal of these evaluations is to
measure the impact of memory fit on a complete end-to-end hardware/software
system running applications on Linux.  We do not run Memtis or HeMem, as they
rely on Intel PEBS~\cite{memtis,hemem}, and event sampling on RISC-V remains an
open problem~\cite{riscv-pebs}.

A Small BOOM core, running at 18MHz,  cannot generate enough traffic to
saturate the FPGA host's HBM or introduce significant queueing at the HBM
interface. To approximate the performance of different memories that a normal
core would observe, we limit available memory bandwidth with FireSim's memory
timing model~\cite{firesim}. The system models both DRAM and PCM, setting PCM's
write bandwidth to 10\% of DRAM's.~\cite{zilberberg13pcm} Prior work shows that
multicore servers can reach PCM's write bandwidth limits directly, saturating
bandwidth with as few as four threads~\cite{optane}. The goal of this setup is
not to precisely recreate the performance of PCM-based memory technologies, but
rather to evaluate the complete hardware/software memory fit architecture when
there are two memories with distinct tradeoffs.  Both DRAM and PCM have a
minimum delay of 16 cycles.

We run workloads with the same memory capacities and migration limits on a
kernel using memory fit and the same kernel using TPP.  They share the capacity
controller and migration mechanism, differing only in how they rank pages. All
pages are 4\,KiB. For capacity sweeps, we report DRAM capacity as a percentage
of the workload's memory footprint.  Exploring memory fit for large (2MB) and
huge (1GB) pages is future work.  We measure the total CPU cycles needed to
complete an application workload. This measurement includes controller and
migration costs in online runs.

Because we seek to evaluate how well memory fit aids an operating system in
page placement, rather than the absolute performance tradeoffs of simulated
memories, we measure each setup relative to the worst case and optimal page
placement. We report this value as the {\it cycle gap recovery}: what
percentage of the difference between worst case and best case performance does
a configuration recover? 100\% means it performs as well as the optimal page
placement and 0\% means it performs as poorly as the worst case placement.
Random placement performance can be complex with heterogeneous
memories, but a write heavy microbenchmark in a 40\% PCM/60\% DRAM
setup would expect to recover 60\% of the performance gap.
In
our configuration, there is approximately a 10\% absolute performance difference
between best case and worst case placements.
Cycle gap recovery
shows how much of the available placement benefit a policy captures.
We also
report direct changes in runtime or throughput.

\subsection{Placement Quality}
\label{sec:extended-gups}

\begin{figure}[t]
  \input{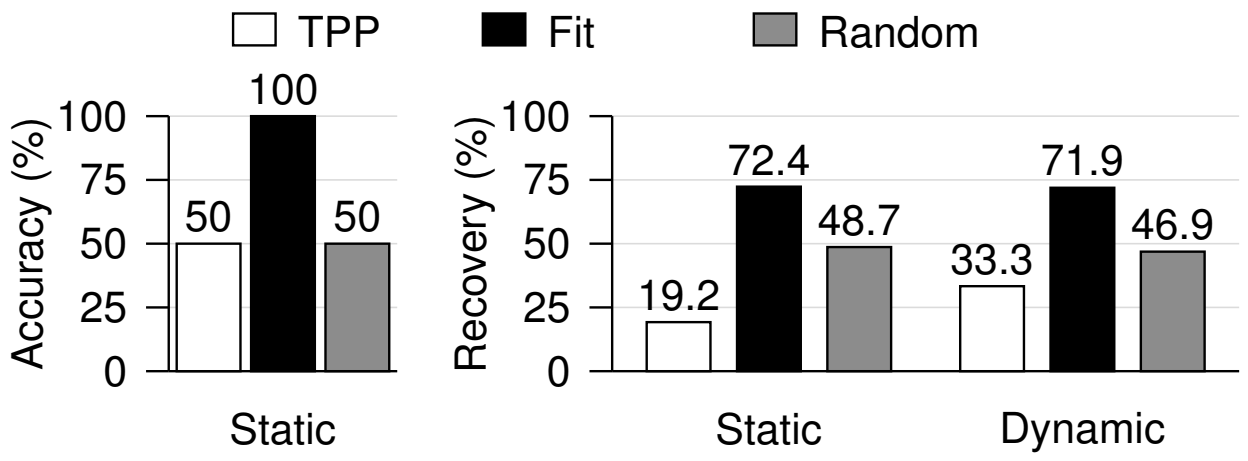}
  \caption{Placement accuracy and cycle-gap recovery of the extended
    GUPS benchmark. TPP places no better than random while memory fit
    has perfect accuracy.}
  \Description{Two black-and-white bar plots. Static placement accuracy is
    100 percent for STM and 50 percent for both TPP and random placement.
    Performance recovery is 72.4 and 71.9 percent for STM, 19.2 and
    33.3 percent for TPP, and 48.7 and 46.9 percent for random placement.}
  \label{fig:microbenchmarks}
\end{figure}

We measure memory fit's ability to identify different access patterns using
an extended GUPS benchmark. The benchmark
has 3 regions of memory: a read-only group of 6,144 pages, a write-only group of
6,144 pages, and a group of 12,288 pages that are not accessed.
DRAM capacity is sufficient to hold one active group. The optimal placement
puts the write-only pages in DRAM, while the worst placement puts
the write-only pages in PCM.
In the
static configuration, the read and write groups do not change during the run.
In the dynamic configuration, which group is read-only and which group is write-only
changes halfway through the run. The dynamic configuration measures how well
a system adapts to sudden changes in access behavior.

Figure~\ref{fig:microbenchmarks}\subref{fig:microbenchmarks:accuracy} shows that STM-based memory fit
places all write-heavy pages in DRAM and read-only pages in PCM, while TPP and
random placement each reach 50\% accuracy.
Figure~\ref{fig:microbenchmarks}\subref{fig:microbenchmarks:recovery} shows that this accuracy allows the
memory fit system to recover 72.4\% and
71.9\% of the optimal-to-worst cycle gap for Static and DYN. TPP recovers
19.2\% and 33.3\%, while random placement recovers 48.7\% and 46.9\%. TPP
performs better in the dynamic setting because its accuracy is so low --
the fact that it estimates wrongly means it benefits from the change.
The difference between accuracy and recovery is significant because on our
single-issue BOOM core, moving a page in software is expensive.
Memory fit distinguishes pages that benefit differently from placement
in DRAM even when their access counts are equal.

\subsection{Cost and Tradeoffs}
\label{sec:eval-comparisons}

\begin{figure}[t]
  \input{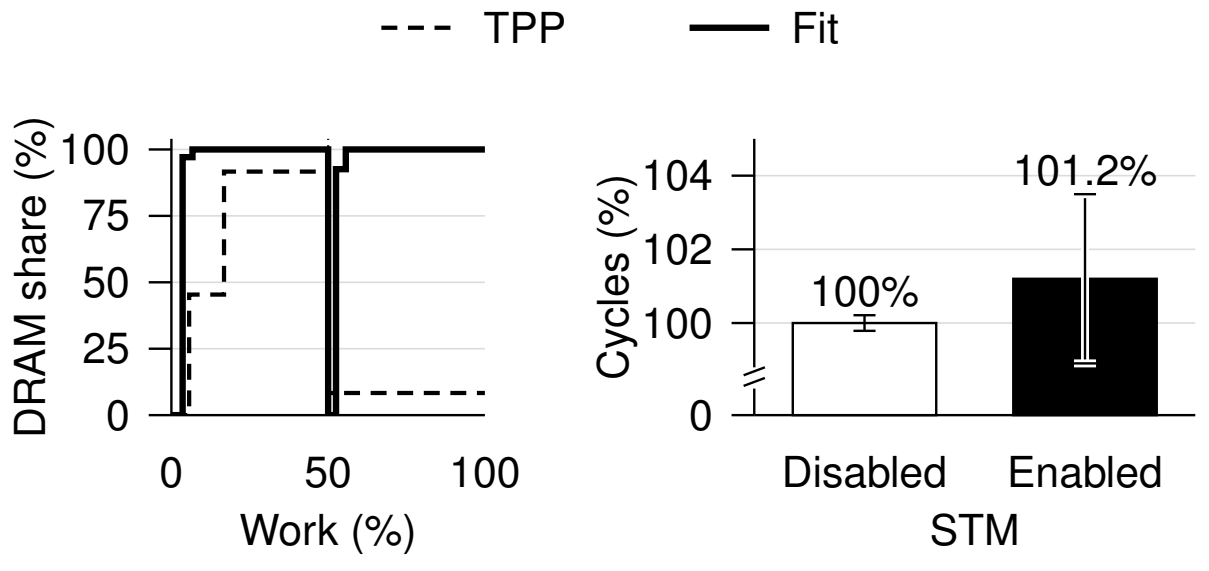}
  \caption{GUPS phase-change adaptation and observation cost. STM restores
    write-heavy placement; observation adds 1.2\% overhead.
    Whiskers show one standard deviation.}
  \Description{Two black-and-white plots. The migration plot shows the
    write-heavy share of DRAM. STM reaches 100 percent in both phases,
    while TPP does not react after the behavior change. With placement fixed,
    enabling STM observation and notification processing adds 1.2 percent
    to execution time.}
  \label{fig:microbenchmark-cost}
\end{figure}

Figure~\ref{fig:microbenchmark-cost}\subref{fig:microbenchmark-migration} shows how page placement changes
over time for TPP and memory fit. Because it observes every main memory access,
memory fit is able to very quickly identify which pages are write-heavy and migrate
them accordingly. TPP, in contrast, takes a long time and is inaccurate.

Figure~\ref{fig:microbenchmark-cost}\subref{fig:microbenchmark-overhead} shows the execution time overhead of
memory fit's notifications and corresponding processing. It measures the
execution time when placement and access patterns are fixed, with memory
fit enabled and disabled. Memory fit's observation and processing add 1.2\%
to execution time.
Appendix~\ref{app:prototype-overhead} provides further details on the measurement
methodology.
In our single-core evaluations, memory fit's software processing runs on the same
core as the workload and directly impacts end-to-end performance. In
multicore systems, we expect this overhead to vanish by assigning
interrupt handling and notification processing to a separate core, as prior
work running on commodity processors does.

\subsection{End-to-End Application Performance}

\begin{figure}[t]
  \centering
  \includegraphics
    {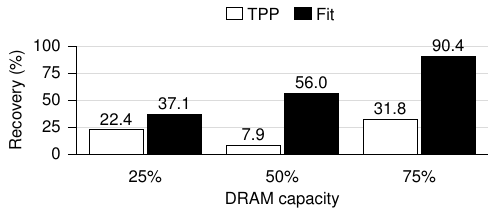}
  \caption{Redis cycle-gap recovery across DRAM capacities.
    STM outperforms TPP at every capacity.}
  \Description{A black-and-white grouped bar plot at DRAM capacities of 25,
    50, and 75 percent. STM recovers 37.1, 56.0, and 90.4 percent;
    TPP recovers 22.4, 7.9, and 31.8 percent.}
  \label{fig:redis-end-to-end}
\end{figure}

\begin{figure}[t]
  \centering
  \includegraphics
    {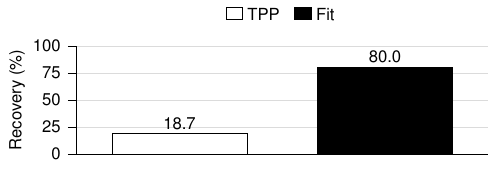}
  \caption{Redis placement replay at 50\% DRAM capacity.
    STM's placements recover 80\% of the cycle gap, versus 19\% for TPP.}
  \Description{Two black-and-white bars show cycle-gap recovery of
    80.0 percent for STM replay and 18.7 percent for TPP replay.}
  \label{fig:redis-replay}
\end{figure}

We measure the end-to-end performance of two complete applications, Redis and
llama.cpp, running with TPP and memory fit. The Redis workload combines two
client access patterns: batched key-value lookups through independent reads that permit
memory-level parallelism and  pointer-based traversal with
per-node updates (e.g., graph processing).
The llama.cpp workload uses the Q8\_0 TinyStories-15M
model with eight contexts sharing its weights. We omit the model's arithmetic
computation to keep the workload from becoming compute-bound on our single-core
prototype. The workload accesses model
weights, activations, and KV caches under the same write-bandwidth asymmetry.

We run Redis across three DRAM capacities to test whether better page
choices improve application performance. Figure~\ref{fig:redis-end-to-end}
normalizes the cycle gap between placing all pages in DRAM (100\%) and
placing all pages in PCM (0\%). At configured DRAM capacities
of 25\%, 50\%, and 75\%, STM recovers 37.1\%, 56.0\%, and 90.4\%, versus TPP's
22.4\%, 7.9\%, and 31.8\%.
Replaying the captured 50\%-capacity placements without controller or migration
activity recovers
80.0\% for STM and 18.7\% for TPP (Figure~\ref{fig:redis-replay}).

\begin{figure}[t]
  \centering
  \includegraphics{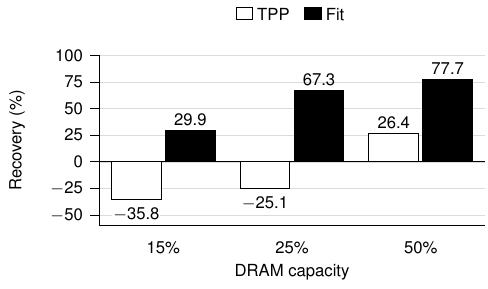}
  \caption{llama.cpp cycle-gap recovery. STM outperforms TPP at all
    three DRAM capacities.}
  \Description{Three pairs of black-and-white bars at DRAM capacities of
    15, 25, and 50 percent. STM recovers 29.9, 67.3, and 77.7 percent;
    TPP recovers minus 35.8, minus 25.1, and 26.4 percent. Negative recovery
    means more cycles than the all-PCM placement.}
  \label{fig:llama-end-to-end}
\end{figure}

For llama.cpp, we measure 24 passes, including initial convergence and all online controller
and migration costs. We use DRAM capacities of 15\%, 25\%, and 50\% to test
whether memory fit retains its advantage as more pages fit in DRAM.
Memory fit recovers 29.9\%, 67.3\%, and 77.7\% of the
all-PCM-to-all-DRAM cycle gap, versus TPP's $-35.8\%$, $-25.1\%$, and
26.4\% (Figure~\ref{fig:llama-end-to-end}).

Interestingly, TPP's excessive migration and access-monitoring overhead outweigh its
placement benefit at smaller capacities. This leads to negative recovery.
TPP performs substantially more page moves than STM and incurs frequent
minor page faults for access sampling, adding significant overhead on the
same core as the workload. Memory fit, in contrast, prioritizes the write-sensitive activation
and KV pages, recovering more performance at every capacity.

\subsection{Other Memories}

\begin{figure}[t]
  \centering
  \includegraphics
    {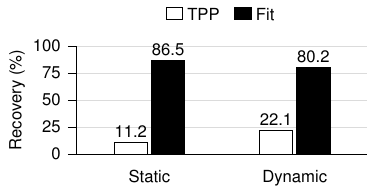}
  \caption{CXL-like cycle-gap recovery. STM outperforms TPP for
    both Static and Dynamic.}
  \Description{Two pairs of black-and-white bars. STM recovers 86.5 percent
    of the static best-to-worst cycle gap and 80.2 percent of the dynamic gap;
    TPP recovers 11.2 and 22.1 percent, respectively.}
  \label{fig:cxl-hotcold}
\end{figure}

We evaluate whether memory fit can incorporate other memory technologies,
such as a higher-latency CXL device.
To emulate a CXL-like device, we add latency to one memory, load a memory
fit model trained for this memory, and run the
extended GUPS in Section~\ref{sec:extended-gups} with added hot/cold separation. With memory fit,
STM recovers 86.5\% and 80.2\%
of performance, compared with 11.2\% and
22.1\% for TPP (Figure~\ref{fig:cxl-hotcold}).
This experiment demonstrates the retargetability of the memory fit interface
across different memory-system setups.

\subsection{Implementation Overhead}
\label{sec:implementation-overhead}

The memory fit system uses
10,414 LUTs, 7,572 flip-flops, and 13 DSP blocks for STM and its request
monitor. These totals include the MMIO,
delta-cache, and metadata-commit logic, with no block RAM or UltraRAM.
Altogether, memory fit, STM, and  the request monitor add 15.1\% LUTs and 15.9\%
flip-flops to a Small BOOM tile.
Against a reference budget of 350K--500K LUT
equivalents per large server core, memory fit would be 2.1--3.0\% of one core's
LUT budget.  Memory fit is shared uncore hardware: its cost amortizes across cores using
the observed memory path.
Shared across 96 cores, it uses 0.022--0.031\% of their combined
LUT budget. Appendix~\ref{app:lut-comparison} gives the reference budgets,
sharing calculation, and replication assumptions.

Furthermore, the current RTL prioritizes functional validation and timing closure.
An implementation optimized for area has substantial room to reduce both
LUT and flip-flop use. Section~\ref{sec:eval-comparisons} measures the runtime
cost of maintaining the memory fit records.

\subsection{Composition with Other Placement Techniques}
\label{sec:eval-composition}

\begin{figure}[t]
  \centering
  \includegraphics{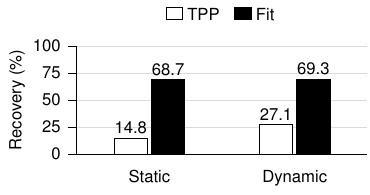}
  \caption{Extended GUPS performance with Colloid and per-page MLP ranking.}
  \Description{STM and TPP bars for Static and Dynamic. STM recovers 68.7 and
    69.3 percent of the reference cycle gap, versus 14.8 and 27.1 percent
    for TPP. Both policies use Colloid and per-page MLP ranking.}
  \label{fig:microbenchmark-composition}
\end{figure}

Deciding where to place pages is only part of managing heterogeneous memories: prior
work has shown that, due to bandwidth, carefully controlling when to move pages is
also extremely important. For memory fit to be useful, it needs to be able to compose
with these more advanced page movement policies.  We measure how a memory fit
system performs when combined with the state-of-the-art techniques,
Colloid~\cite{colloid} and ALTO's memory level parallelism (MLP)
ranking.~\cite{beyondhotness} This extended GUPS variant uses three
4,096-page active groups with parallel reads, parallel writes, and
serialized read-modify-writes, plus a 12,288 cold page
region.
DRAM is sized to be able to hold two active groups. The
dynamic configuration swaps the parallel read and serialized accesses
halfway through the experiment.
Figure~\ref{fig:microbenchmark-composition} shows that memory fit recovers
68.7\% and 69.3\% of performance, compared with 14.8\% and 27.1\% for TPP. Memory
fit improves workload throughput over TPP by 13.2\% in Static and 10.3\% in DYN, showing
that memory fit retains its placement advantage alongside these complementary
techniques.

\subsection{Overhead Reduction of Hardware Filters}
\label{sec:hardware-ablations}

\begin{figure}[t]
  \input{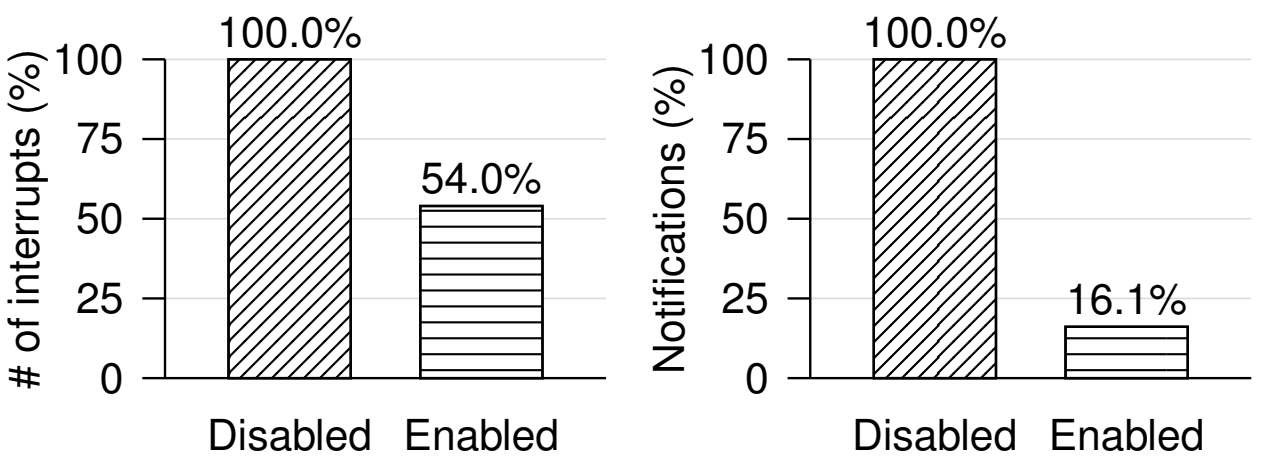}
  \caption{Coalescing reduces interrupts by 46.0\%; low-pass
    filtering reduces decision notifications by 83.9\%.}
  \Description{Two bar plots label their bars Disabled and Enabled. Disabled
    bars have diagonal hatching; enabled bars have horizontal hatching. Both
    use white backgrounds and black outlines.
    Each disabled baseline is 100 percent. With coalescing enabled,
    interrupts are 54.0 percent of baseline. With low-pass filtering
    enabled, decision notifications are 16.1 percent of baseline.}
  \label{fig:notification-ablations}
\end{figure}

Much of the implementation complexity of memory fit and STM focuses on
reducing the number of updates the OS must handle. To measure the impact
of these features, we measure performance on the extended-GUPS workload with fixed
placement and migration disabled. Coalescing reduces ring publications and
interrupts by 46.0\%
(Figure~\ref{fig:notification-ablations}\subref{fig:notification-coalescing}).
Low-pass filtering reduces decision notifications by 83.9\%, with coalescing
enabled in both runs
(Figure~\ref{fig:notification-ablations}\subref{fig:notification-low-pass}).
A before-and-after comparison of notification and controller refinements in
the llama.cpp workload reduces late-run cycle overhead relative to all-fast
from 3.7\% to 0.26\%, without sacrificing STM's placement quality.

\section{Related Work}
Soft trace matching relates to counter- and learning-based page placement,
branch prediction, and context modeling.
\label{sec:related}

\textbf{Counter-based Page Placement} 
Page placement research for tiered memory and heterogeneous memory
systems has relied on counters, as commodity processors provide them.
Prior systems use performance counters
and page-fault hints (e.g., Intel PEBS) to estimate page hotness and guide
placement \cite{hemem, tpp, flexmem, memtis, nomad}. Recent work derives
heuristics to capture memory-level parallelism \cite{beyondhotness},
interconnect contention \cite{colloid}, and VM page-allocation conflicts
\cite{vmcxl}. Hardware-managed schemes make similar hotness estimates but at
finer granularity and higher measurement fidelity \cite{m5}. Memory fit differs
by introducing a ``soft'' counter that measures how well an access pattern
matches a memory type's performance characteristics, rather than only access
frequency. 

\textbf{Learning-based Page Placement} 

Learning-based policies range from deep
reinforcement learning for predicting future access frequency \cite{kleio} to
learned page demotion criteria \cite{idt}. Memory fit and STM differ from these
learning-based methods by providing an accurate, efficient model of the
relationship between memory types and access patterns without neural networks
or online training. This permits an efficient implementation that can track
every access to every page.

\textbf{Branch Prediction}
Using memory fit for placement has many conceptual similarities to branch prediction:
they both need to accurately predict highly dynamic behavior with low overhead
and state.
Similar to lightweight dynamic predictors (e.g., pattern history tables and saturating
counters) that exploit history correlations \cite{branchpred, cbp}, soft trace
matching uses compact state to map an observed sequence to a memory fit score.
Unlike dynamic branch predictors, its learned tables are based on the memory type
and fixed at runtime.

\textbf{Context Modeling}
Context Tree Weighting (CTW) organizes conditional symbol counts in a tree and
combines predictions across context depths \cite{ctw}.  Soft trace matching
borrows this tabulation structure for nested access-pattern counts, but it does
not encode or reconstruct data: it is an estimator rather than a lossless compressor.
Accordingly, STM's tables and reuse anchors are trained for a comparative
placement objective, and a root-only table remains a complete model; additional
context rows are refinements when a memory pair benefits from history.

\section{Discussion and Conclusion}
\label{sec:conclusion}

This paper proposes memory fit as an architectural mechanism to support
operating system policies for page placement in systems with heterogeneous
memories.  Soft trace matching implements memory fit as a compact, learned
comparison between an observed page-access sequence and the access
distributions associated with each memory.

The STM implementation described in this paper uses records that encode
read/write and reuse distance; this enables encoding read/write behavior, 
access frequency, and potential parallelism. The core STM algorithm, however,
is more general, and can use other record representations. While the results
in this paper show that these records can capture important properties of 
existing memories, other record representations may improve on these results
or be needed to capture future memory tradeoffs.

Likewise, while STM as implemented in this paper is trained for a specific definition
of "better" - lower completion time - other objective functions could be used,
such as a weighted composite score that balances energy efficiency, performance,
and other metrics like durability and fairness that are desirable in a system.
STM is, therefore, a more general method of learning
"fitness" beyond the specific objective used in this paper, and could be extended to different objectives as needed.

At the same time, while memory fit in this paper uses STM to estimate 
access affinity to different
memories, it is open to other algorithms or estimators.
Whether other algorithms can outperform STM is an interesting 
area of future investigation.

Much of the complexity of the implementation described in this paper is due
to its providing coherent and precise estimates to the operating system. It
may be that relaxing some of these guarantees could reduce overhead without
substantially reducing accuracy. Scaling memory fit to multi-chiplet, high core-count
processors, or multi-socket systems might require such relaxations.

A critical property of memory fit scores is that they are comparable across pages and memories.
Hardware maintains the
page-local signal off the memory request path, while Linux ranks feasible moves and
performs migration through its existing page-management path.  The
implementation stores eight bytes per tracked 4\,KiB page in a separate
PFN-indexed table and requires no PTE or \texttt{struct page} change.
The FPGA prototype improves placement over TPP across read/write-asymmetric
and CXL-like memory setups, with 1.2\% measured observation and notification
overhead at fixed placement. These results suggest, as memory grows increasingly
heterogeneous and complex, operating systems might increasingly benefit from  a hardware interface
that separates measurement mechanisms and operating-system placement policy.

\clearpage
\appendix
\section{Formal Memory Fit Model and Placement Bound}
\label{app:formal}

This appendix formalizes the training target and steady-state placement bound used
in Sections~\ref{sec:fit-semantics} and~\ref{sec:placement-loop}.

\subsection{Counterfactual Training Target}
\label{app:calibration}

Let $\mathcal{M}$ be the set of available memories, with $m$ denoting one
memory.  A workload instance $x$ specifies fixed work and the target pages
whose placement is varied.  Let $C_m(x)>0$ be its completion time in cycles
when those pages reside in $m\in\mathcal{M}$.  For distinct memories
$a,b\in\mathcal{M}$, define the dimensionless training target $y_{a,b}(x)$
from paired counterfactual runs.  Let $A>0$ be a fixed training-scale
constant, and let $\operatorname{clip}(v,-1,1)$ clamp its scalar input $v$
to the interval $[-1,1]$.  Throughout this appendix, $\log$ denotes the
base-two logarithm, so logarithmic losses are measured in bits.

\begin{equation}
  y_{a,b}(x)=\operatorname{clip}\left(
    \frac{\log(C_b(x)/C_a(x))}{A},-1,1\right),
  \label{eq:cycle-advantage}
\end{equation}

Only target-page placement changes between the paired runs.  Positive
$y_{a,b}(x)$ means that $a$ is faster.  For each observed page $p$, soft trace
matching maintains a loss $S_m(p)$ for each memory $m$; lower is better.
Equation~\ref{eq:page-loss} defines this page-level average.  The pairwise
signal $\Delta_{a,b}(p)$ is the loss on $b$ minus the loss on $a$:

\begin{equation}
  \Delta_{a,b}(p)=S_b(p)-S_a(p).
  \label{eq:score-gap}
\end{equation}

The sign chooses a memory for one page and the magnitude ranks pages competing
for constrained capacity.  An individual $S_m$ is not a cycle estimate;
training fits the relative memory fit score to
Equation~\ref{eq:cycle-advantage}.

\subsection{Trace Representation and Nested Counts}
\label{app:trace-model}

A rule over several counters can represent a known distinction.  For example,
a designer could count reads and writes separately and assign them different
weights.  That solution already has the central structure of a model: an event
representation, state that tabulates those events, and a rule that maps the
state to a placement preference.  The difficult question is not whether a
counter can be added.  It is how to define and train this model so that
it can be extended when a different memory makes reuse, operation order, or
another observable property important.  STM is such a counter model and makes
these choices explicit.  A single-table STM instance is a set of event counters
with learned weights.  Additional tables condition those counts on recent page
history when the memory requires it (Section~\ref{sec:stm-computation}).

\begin{figure*}[t]
  \centering
  \includegraphics[width=\textwidth]{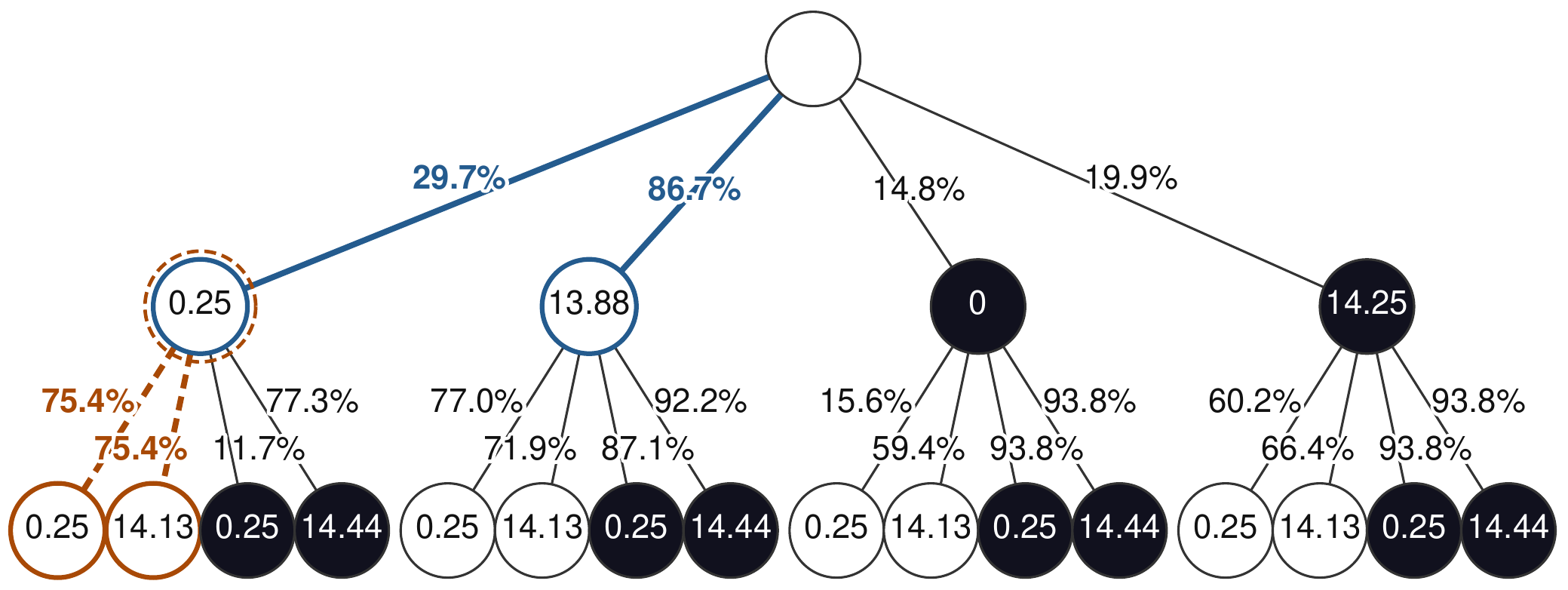}
  \caption{STM computation on a learned tree for a synthetic CXL-like memory
  profile. Hollow and filled nodes denote read and write events. Node values
  are normalized reuse anchors; edge labels are learned match weights shown
  as percentages. Labels are rounded for display. A previous short-reuse read
  selects the leftmost context, outlined in dashed orange. A current read
  midway between the root's read anchors (reuse $\approx 7.06$) interpolates
  the solid blue branches' 29.7\% and 86.7\% weights to obtain 58.2\%.
  The dashed orange branches both give 75.4\% in the selected context.
  Mixing the root and conditional outputs with weights 75\% and 25\%
  yields 62.5\%. Its negative log, approximately 0.68 bits before hardware
  log approximation, updates the page's decaying memory fit score.}
  \Description{A complete four-way tree has a root, four intermediate nodes,
  and sixteen leaves, using learned CXL-like match weights and reuse anchors.
  A previous read below the learned short/long threshold selects the leftmost
  context. A current read at reuse 7.0625 interpolates equally between the
  root's read anchors, 0.25 and 13.875. The root gives approximately 58.2
  percent, and the selected context gives 75.4 percent. Their 75/25 mixture
  gives 62.5 percent, whose negative log is approximately 0.68 bits.}
  \label{fig:stm-inference}
\end{figure*}

Figure~\ref{fig:stm-inference} expands the tables in
Section~\ref{sec:stm-computation} into a complete trace tree. Each set of four
children represents read and write events at short- and long-reuse anchors.
Outgoing edge weights form the corresponding table: the root gives marginal
weights, while an intermediate node gives weights conditioned on that
preceding event.

The caption walks through one read observation using the highlighted
branches. STM repeats this computation with each memory's own tables.

For a page $p$, index its observations in temporal order by $j$.  The symbol
$z_j$ records the operation type $o_j\in\{\mathrm{read},\mathrm{write}\}$
and normalized reuse distance $r_j$.  Let $A_j$ be the global count of
observed memory accesses at observation $j$, and let $A_{j-1}(p)$ be that
count at the preceding observation of the same page.  These access counts
are distinct from the training-scale constant $A$ in
Equation~\ref{eq:cycle-advantage}.  Let $\widehat{W}>0$ be the estimated
number of distinct pages in the current working set.  Then

\begin{equation}
  z_j=(o_j,r_j), \qquad
  r_j=\frac{A_j-A_{j-1}(p)}{\widehat{W}},
  \label{eq:trace-symbol}
\end{equation}

Thus $r_j$ expresses inter-access delay relative to the working-set estimate.
Let $h_j$ denote the history preceding $z_j$ on the same page.  Symbols
preceding the first indexed observation belong to the initial history.
For a nonnegative integer $k$, the $k$th-order model $Q_{m,k}$ for memory $m$
predicts the current symbol using only the previous $k$ symbols:

\begin{equation}
  Q_{m,k}(z_j\mid h_j)=
  Q_{m,k}(z_j\mid z_{j-k},\ldots,z_{j-1}).
  \label{eq:markov-model}
\end{equation}

Here the vertical bar denotes conditioning on a history.  Order $k=0$ uses
an empty history and gives the root model's marginal prediction.
Short contexts preserve evidence but omit correlations; long contexts
separate behavior while dividing observations among more entries.  Soft trace
matching tabulates suffix contexts in a tree, replaces exact delay matching
with interpolation between learned anchors, and mixes the levels.

Let the nonnegative integer $K$ be the maximum history length, so the model
combines orders $k=0,\ldots,K$.  The two-table implementation uses $K=1$: the root has no
history, and the conditional table uses one preceding observation.
Let $\phi$ map a preceding symbol to its coarse context class: read or write
and short or long reuse, using the learned short/long-reuse boundary.
The encoded suffix of length $k$ is
$h_j^{(k)}=(\phi(z_{j-k}),\ldots,\phi(z_{j-1}))$, with $h_j^{(0)}$ empty.
For each memory, let $\lambda_k\geq0$ be the learned mixture weight for
order $k$; the memory dependence of these weights and the context encoding
is implicit.  The table output $Q_{m,k}(z_j\mid h_j^{(k)})$ is the learned
match value for the current symbol after reuse-anchor interpolation in
that context.  The combined match value $Q_m(z_j\mid h_j)$ is

\begin{equation}
  Q_m(z_j\mid h_j)=\sum_{k=0}^{K}\lambda_k
    Q_{m,k}(z_j\mid h_j^{(k)}),
  \qquad \sum_{k=0}^{K}\lambda_k=1.
  \label{eq:stm-mixture}
\end{equation}

The event loss $\ell_m(z_j)=-\log Q_m(z_j\mid h_j)$ measures the cost of
observation $j$ under memory $m$; its history dependence is implicit in this
notation.  Match values are positive, so this logarithm is defined.
Let $n_p>0$ be the number of observations of page $p$ included in its score.
Their average loss is

\begin{equation}
  S_m(p)=\frac{1}{n_p}\sum_{j=1}^{n_p}\ell_m(z_j).
  \label{eq:page-loss}
\end{equation}

The implementation uses a decaying average so new phases replace old
evidence.  To express the arithmetic average as counts, let $\mathcal{H}_K$
be the set of encoded contexts $h$ of length $K$; each context supplies the
suffixes used by the mixture.  At $K=0$, this set contains only the empty
context.  Let $\mathcal{U}$ be the set of represented current-event classes.
An event class $u\in\mathcal{U}$ includes the operation and represented
normalized reuse value, so a fixed pair $(h,u)$ determines the interpolated
match value.  This current-event representation retains the reuse information
used for interpolation, whereas $\phi$ stores only a coarse class for history.

For $h\in\mathcal{H}_K$ and $u\in\mathcal{U}$, let $N_p(h,u)$ count the
observations of page $p$ with that context and event class.
The normalized count $\bar N_p(h,u)=N_p(h,u)/n_p$ is their fraction of all
$n_p$ observations.  For fixed learned tables, define the loss associated
with this pair as $c_m(h,u)=-\log Q_m(u\mid h)$, where $Q_m(u\mid h)$ is
the combined match value from Equation~\ref{eq:stm-mixture}.
Write $S_m^{(K)}(p)$ for the same page loss as $S_m(p)$ with maximum history
length $K$ made explicit.  Grouping identical $(h,u)$ terms in
Equation~\ref{eq:page-loss} gives

\begin{equation}
  S_m^{(K)}(p)=
    \sum_{h\in\mathcal{H}_K}\sum_{u\in\mathcal{U}}
      \bar N_p(h,u)c_m(h,u).
  \label{eq:nested-counting}
\end{equation}

The model is therefore additive over normalized, interpretable per-page
counts.  At $K=0$, it is a learned weighted sum of marginal counters.  Setting
$\lambda_0=1$ recovers that root-only model exactly; nonzero deeper weights
condition the same counter basis on history.

For the probabilistic interpretation, assume that $Q_m$ is an exact
predictive distribution.  Let $n=n_p$ and $z_{1:n}=(z_1,\ldots,z_n)$ denote
the page's sequence of $n$ observations.  Conditional on a fixed initial
history, the model assigns this sequence probability
$Q_m(z_{1:n})=\prod_{j=1}^{n}Q_m(z_j\mid h_j)$.
In the next equation, $S_a$ and $S_b$ abbreviate $S_a(p)$ and $S_b(p)$ for
this same trace.  Subtracting their summed negative log probabilities gives

\begin{equation}
  n(S_b-S_a)=\log\frac{Q_a(z_{1:n})}{Q_b(z_{1:n})},
  \label{eq:likelihood-ratio}
\end{equation}

so the gap accumulates log-likelihood-ratio evidence.
Let $P$ be the page's true distribution over traces of length $n$, with
$Q_m$ defined on the same trace space.  The expectation operator
$\mathbb{E}_P$ averages over traces drawn from $P$; $H(P)$ is its Shannon
entropy, and $D_{\mathrm{KL}}(P\Vert Q_m)$ is the Kullback--Leibler
divergence from $P$ to $Q_m$.  Expected log loss also satisfies
$\mathbb{E}_{P}[-\log Q_m]=H(P)+D_{\mathrm{KL}}(P\Vert Q_m)$; the lowest-loss
model is closest to the page's observed distribution
\cite{cover06information}.  Classical CTW bounds redundancy for bounded-
memory sources~\cite{ctw}.  Soft trace matching borrows its mixed-order table
organization but learns soft tables for placement, so it does not claim that
universal-coding bound.

\subsection{Steady-State Bipartite Matching Bound}
\label{app:matching}

For a frozen placement snapshot, let pages be the left vertices of a weighted
bipartite graph and capacity slots, labeled by memory, be the right vertices.
Each slot holds one page.  In this subsection, $u$ denotes a capacity slot,
rather than an event class.  The edge $(p,u)$ represents assigning page $p$
to slot $u$, and its weight $w(p,u)\geq0$ is the learned benefit of assigning
that page to the slot's memory.  A matching $M$ is a set of assignment edges
that share no endpoints: each page occupies at most one slot, and each slot
holds at most one page.  A complete assignment covers every page.
A per-page shift can make weights nonnegative without changing the ordering
of complete assignments.  The operator $\arg\max$ selects a feasible matching
with greatest total weight; $M^*$ denotes such an optimal matching.
The global placement is

\begin{equation}
  M^*=\arg\max_{M\ \mathrm{feasible}}
       \sum_{(p,u)\in M}w(p,u).
  \label{eq:placement-matching}
\end{equation}

For any matching $M$, write $w(M)=\sum_{(p,u)\in M}w(p,u)$ for its total
weight.  Let $M_G$ be the greedy matching formed by processing edges in
nonincreasing weight and accepting an edge when both endpoints are free.
Then

\begin{equation}
  w(M_G)\geq\tfrac{1}{2}w(M^*).
  \label{eq:matching-half}
\end{equation}

To see this, charge every optimum edge rejected by a greedy edge to the first
greedy edge sharing one of its endpoints.  A greedy edge receives at most two
charges, one per endpoint, and it is no lighter than either charged edge when
selected.  Summing the charges proves Equation~\ref{eq:matching-half}
\cite{preis99matching}.  The result applies to a complete, fixed edge set.
Bounded queues, controller decisions, and changing memory fit scores restrict the online
driver to an admitted subgraph.  We invoke the guarantee only after its
candidate set and preferences become steady.

The driver's swap rule gives a separate local guarantee.  For a fixed set of
represented candidates and fixed memory fit scores, it accepts a move into an
empty slot only when that move lowers the page's memory fit score.  When no
slot is empty, it accepts an incoming--resident swap only when the pair lowers
the sum of their memory fit scores.  Every successful step therefore strictly
improves the current assignment.  Because the number of page assignments is
finite, the process stops at a placement where no represented one-page move or
two-page swap can improve the objective.  This is a local minimum over the
candidate neighborhood retained by the bounded heaps.  Changing scores or an
omitted candidate falls outside this steady-state claim.

\section{Full Implementation}
\label{app:implementation-details}

We implement soft trace matching as a complete FireSim target and run it on an
AWS F2 FPGA.  The target contains one Small BOOMv3 core, a one-wide
out-of-order RISC-V processor, running Linux 6.6.  Its L1 instruction and data
caches are each 16\,KiB and four-way set associative.  A modeled 4\,MiB,
eight-way last-level cache (LLC) fronts two modeled physical memories with 16\,GiB of
total addressable capacity.  Both use a quad-rank, FR-FCFS timing model, and
Linux sees them as two contiguous 8\,GiB NUMA nodes.
The target core and memory bus operate at 1\,GHz, while the remaining target
buses and STM operate at 500\,MHz.  Because FireSim executes the target at a
lower FPGA-host frequency, the routed image itself runs at 18\,MHz.

STM attaches at the output of the LLC.  A TileLink adapter sends each
post-cache request header toward a separate observation path while forwarding
the demand request unchanged.  If the observation path is full, the adapter
drops the header instead of applying back pressure to the demand request.  Thus
STM sees only traffic that reaches memory and cannot add a pipeline stage to a
cache hit.  A separate TileLink client issues
coherent accesses to the PFN table and event ring without involving the core
pipeline.  MMIO carries only configuration and status.  The following
subsections trace an observation from request capture through model evaluation,
metadata commit, and Linux notification before reporting the cost of the full
implementation.

\subsection{Observation and Delta-Commit Pipeline}
\label{app:delta-commit}

At the LLC output, the adapter fires each memory-request header into a private
ingress queue.  If the queue is full, the adapter drops the header and continues
serving the demand request.  Thus, the demand request does not wait for the
queue to drain, for STM computation, or for a metadata access to complete.  The
queue also transfers accepted headers from the memory-bus clock domain into
STM's clock domain.  After this transfer, address filters discard
requests outside the two tracked memories as well as accesses to the state
table and event ring.  Excluding these metadata accesses prevents STM from
observing work that it generated itself.  The adapter can receive one header
per memory-bus cycle, and the STM pipeline can consume one accepted observation
per STM cycle.  The ingress queue absorbs short differences between these
rates.  Longer bursts may fill the queue and cause later observations to be
dropped rather than slowing the memory-access path.

To avoid a table access for every observation, STM uses a 64-entry, four-way
delta cache.  Each entry holds the changes produced for one page since that
page was last committed to the PFN table.  A hit updates only this on-chip
copy.  On a miss, hardware
starts a new entry at zero instead of first reading the page's PFN record from
memory.  This choice prevents a newly observed page from immediately creating
a dependent metadata read.  When an entry is evicted, a bounded commit queue
sends its accumulated update to the PFN table.  Repeated accesses to a resident
page therefore produce one table read and write at commit rather than one
metadata operation per observation.

Because evictions access memory-backed metadata, the commit queue
also controls how much traffic STM can generate.  If coherent writes fall
behind, its occupancy lowers the probability that a new page evicts an
existing entry.  Pages already in the cache continue to
accumulate updates, but evictions and their associated table accesses fall
as pressure rises.  STM therefore responds to pressure on its own metadata
path without attempting to estimate total memory-bus utilization.

Starting an entry at zero is possible because hardware can summarize a
sequence of memory fit updates as one affine transform per memory.  One term
records how much of the prior memory fit score remains after exponential
decay.  The other records the evidence accumulated during the cache residency.
At commit,
hardware reads the memory-backed record once, applies both terms, and writes the
result.  This non-atomic read-modify-write sequence produces the same decaying
average as applying each observation
in order, even though hardware did not read the PFN table on the initial cache
miss.  The acknowledged write is therefore the commit point.  The merge engine
serializes commits to the same PFN before it publishes a notification.

\subsection{STM Computation and Approximations}
\label{app:quantized-updates}

The STM datapath begins by combining operation type (read or write) with a
normalized reuse distance measured relative to the current working set.  Raw
reuse distance alone is difficult to interpret because the same distance represents
frequent reuse in a large working set and infrequent reuse in a small one.  An
eight-bit timestamp advances after a fixed number of observed requests.  For a
page that remains in the delta cache, subtracting its previous timestamp
estimates how many memory-side accesses occurred between references.

However, the first observation after a delta-cache installation has no
page-local timestamp because installation deliberately avoids reading the PFN
table.  Hardware uses a global reuse estimate for this observation.  At commit,
hardware compares the timestamp in the old PFN record with the first timestamp
stored in the new delta entry.  These samples update the global estimate used
by later installations.  Subsequent observations use the page-local timestamp
normally.  This approximation preserves the no-read installation path while
providing a reasonable reuse value at a cache-residency boundary.

In parallel, a 16-register HyperLogLog estimator~\cite{hyperloglog} estimates
the number of distinct pages in the current working set.  Dividing reuse
distance by this estimate places reuse on a working-set-relative scale and allows the same
learned anchors to apply across different working-set sizes.  STM uses the
normalized distance only to select and interpolate between nearby probability
entries.  It does not expose the timestamp or working-set estimate as a
performance measurement.  Their limited role permits a wrapping timestamp
and an intentionally coarse cardinality estimate.

The evaluated tree instantiates two table levels.  The root uses the current
event alone, while the conditional level also selects on the preceding coarse
read/write and short/long-reuse event.  This depth-two organization implements
the prototype choice described in Section~\ref{sec:stm-computation}.  It is not a
restriction on the event representation or tree depth of STM itself.

Figure~\ref{fig:stm-inference} illustrates how STM evaluates one observation.
The root table first estimates the probability of the current event without
using prior history.  A second table refines that estimate based on the
preceding coarse read/write and short/long-reuse event.  Each table stores learned
probabilities at short- and long-reuse anchors.  Hardware interpolates between
these anchors at the observed reuse distance, combines the root and
conditional probabilities, and approximates the negative log of the combined
probability.  An event that is common under a memory's model produces a small
update.  An event that the model considers unlikely produces a larger update.
The two memory models then perform this computation in parallel using
different learned tables, and each result updates one eight-bit,
exponentially weighted memory fit score.  An event that does not distinguish the memories moves both
scores similarly.  A sequence that one model explains better gradually
separates the two scores.  Because a lower memory fit score is better, the
model that consistently assigns higher probability becomes preferred.

Two randomized mechanisms preserve information that would otherwise require
wider state.  First, the affine transform retains fractional precision, but
the memory-backed record holds only eight bits per memory fit score.  Truncating a
sub-unit change would discard it and prevent repeated small updates from
accumulating.  Stochastic rounding instead chooses between the two adjacent
integer values according to the fractional part.  In expectation, this
produces the same update as full-precision arithmetic.  Both memories share
the same random draw so rounding cannot introduce an artificial ordering
reversal.

Second, randomized tickets act as a low-pass filter when the two memory fit
scores are close.  Small updates can repeatedly reverse their order.
Publishing every reversal would make Linux follow noise.  After a configurable
sampling step, hardware assigns each observed preference a geometrically
distributed ticket strength.  Each additional level is half as likely as the
preceding one.  The delta cache retains only the strongest four-bit ticket seen
for each preferred memory.  At merge, hardware compares that ticket with the
page's three-bit hysteresis exponent, which is its current stability level.  At level
three, for example, one in eight sampled observations passes on average.  A
confirmed reversal raises the level and halves the chance of another immediate
reversal.  Sustained agreement lowers
the level.  Thus the page's recent instability determines how strongly
reversals are filtered.  Keeping the maximum ticket is sufficient because a
stronger ticket also passes every lower stability level.  Hardware can
therefore postpone the comparison until commit without fetching the memory-backed
stability state.

The optional memory-level-parallelism (MLP) extension counts outstanding
memory reads and maps the count into four classes: one, two, three to four, or
more than four.  A single observation can be noisy, so the delta cache commits
a new class only after repeated observations agree.  Linux may use the class
to refine the order of candidates with similar memory fit benefits.  The MLP
class remains a separate ranking input and does not alter STM computation or
either memory fit score.

\subsection{Training and Model Quantization}
\label{app:training-overview}

Training specializes the STM computation pipeline to a specific memory setup.
For each fixed-size window, the simulator runs the same trace multiple times,
changing only the memory that contains the target pages.  It records the
completion time and post-cache event stream for each run.  The paired
completion times identify the better memory and the relative magnitude of its
advantage.  We implement the STM computation pipeline as a differentiable
function so automatic-differentiation frameworks can adjust the probability
tables, reuse anchors, short/long-reuse boundary, and mixture weights.  The
training objective makes the lower average memory fit score identify the
faster memory and makes the memory fit gap track its relative benefit.

To generate these fixed-size windows, we use synthetic stimuli that cover
read/write mixtures, reuse scales, conditional contexts, locality,
concurrency, warm and cold state, and phase changes.  The stimuli need not
represent an application mix because their purpose is to cover access patterns
that may produce different performance responses from the two memories, not
to tune the trees to a specific workload.  Once training is complete, we
quantize each learned tree weight and probability to the nearest eight-bit
value supported by the hardware tables.

The prototype compiles the learned constants into its hardware image.  Because
the tables and weights are small, an implementation could instead expose them
as MMIO-programmable registers and load a trained model at boot.  This
choice does not change the pipeline or shared record format.

\subsection{Coherent Interface and Linux Driver}
\label{app:coherent-protocol}

Figure~\ref{fig:record-layout} shows the record used by this protocol.
Bits 63:48 are reserved.  Bits 15:12 hold context, bit 11 records an
outstanding notification, bits 10:9 encode the stable memory, and bit 8
validates MLP.  Bits 7:6 hold the MLP class, 5:4 the agreement count, 3:1
the hysteresis exponent, and bit 0 validates the stable direction.

At probe time, the driver allocates a DMA-coherent state table and a
4,096-entry event ring.  It programs their physical bases, the two tracked
address ranges, and policy parameters through MMIO.  Observation begins only after both shared structures
are initialized.  MMIO is therefore used for low-rate setup and status, while
memory fit records and notifications move through coherent memory.

Hardware writes a complete PFN record before publishing the corresponding PFN
to the event ring.  An outstanding-notification bit prevents a page from
occupying more than one unread ring entry.  Hardware atomically commits the
new record while setting this bit.  If the bit was already set, hardware omits
the duplicate ring entry because the existing entry will cause Linux to read
the newest record.  When Linux drains that entry, it atomically obtains the
current record while clearing the bit.  A concurrent hardware update is
therefore included in the record Linux receives or observes the cleared bit
and publishes a later notification.  This protocol coalesces repeated changes
without losing the final state.

As notifications accumulate, the ring raises an interrupt when it reaches a
programmable watermark or when a timeout expires with unread entries.  The
watermark batches notifications under load, while the timeout prevents an isolated notification from waiting
indefinitely.  The interrupt worker drains the available PFNs and advances the
software-owned ring head.  Each PFN is a prompt to inspect the state table, not
a snapshot of the decision that originally produced the event.

After draining the ring, the worker rereads the current record and computes
the benefit of moving the page from its present memory to the preferred
memory.  It inserts eligible pages into one bounded max heap for each source memory.  Each heap deduplicates
PFNs and orders pages by memory fit benefit, allowing a later notification to
update an existing candidate rather than create another copy.  The migration
worker pops the strongest candidate, resolves its current PFN state, and
discards the candidate if the page or memory fit record has changed.  It then
migrates bounded batches through Linux's existing page-migration path.

To make those migration decisions executable, the two tracked ranges use
dedicated movable pageblocks.  This keeps the managed pages eligible for
Linux's existing migration path and avoids mixing them with pages that the kernel cannot move.  Because the state
table is indexed by PFN, moving a page also changes the address of its metadata.
Before migration, the driver commits any resident page delta.  After a
successful copy, it transfers the 64-bit record to the destination PFN and
clears the source.  A failed copy retains the source record.  The driver
serializes this transfer with hardware commits so the page has one
authoritative record before and after migration.

\subsection{Server-Scale LUT Comparison}
\label{app:lut-comparison}

The post-route report measures 10,414 LUTs and 7,572 flip-flops for STM and its
request monitor, compared with 68,805 LUTs and 47,532 flip-flops for the Small
BOOM tile. The respective ratios are 15.1\% and 15.9\%. These compare FPGA
resources in the same image. STM also uses 13 DSP blocks, with no block RAM
or UltraRAM.

For a server-scale comparison, we hold the monitor at 10,414 LUTs and use an
assumed budget of 350,000--500,000 LUT equivalents per large core. Dividing
the monitor's LUT count by these budgets gives 2.1--3.0\% of one core.
The core budgets are reference assumptions, rather than measured FPGA
mappings of a commercial server core. A 96-core processor, such as the
EPYC~9654~\cite{amd9004}, has a combined reference budget of
33.6M--48M LUT equivalents. One shared monitor uses 0.022--0.031\% of that
budget. These percentages compare LUT resources; flip-flops and DSP blocks
are reported separately, and no conversion to transistor count or silicon
area is applied.

STM is an uncore component shared by the cores whose requests reach its
observation path. Its replication follows the number of observation paths
and the throughput each instance must sustain. For $N$ monitor instances,
the total LUT cost is $N\times10{,}414$, and the percentage grows in
proportion. The one-monitor comparison assumes that one instance can service
the observed request stream. It holds the prototype's resource count fixed,
leaving substantial opportunities for an implementation optimized for area
to reduce LUT and flip-flop use.

\subsection{Training Objective and Quantization}
\label{app:training-details}

Training holds work and background placement fixed while changing the memory
containing the target pages. The paired completion times supply the signed
relative advantage in Equation~\ref{eq:cycle-advantage}. A differentiable STM
model consumes the resulting post-cache event stream, maintaining separate
scores and context for each target page. Its prediction is the average
score difference across target pages, weighted by their saturating observation
counts and divided by 255 to match the target range.

The objective combines two terms. A smooth-$L_1$ loss fits the magnitude of the
normalized advantage, including near-ties. A pairwise ranking loss preserves
the ordering of examples from the same workload family, phase, and memory
profile. The ranking target uses the unclipped advantage so clipping the
magnitude target does not erase the order of large benefits. These losses
jointly fit the probability entries, reuse anchors, short/long boundary, and
mixture weights. The trees' context organization is fixed; training learns
their contents rather than searching for a different topology.

The differentiable model includes quantization through straight-through
estimators and models the integer probability lookup and approximate
negative-log datapath. Export rounds and bounds the probability entries,
anchors, threshold, and mixture weights to the hardware representation, then
evaluates the integer model. The FPGA experiments use trained constants
compiled into the RTL. Figure~\ref{fig:stm-inference} illustrates the
computation using a learned tree for the synthetic CXL-like training profile.

Memory fit averages event loss; it is not an access count multiplied by a
per-access latency. Frequent reuse can increase the measured advantage of a
memory and therefore the learned score gap, but access frequency alone does
not force this relationship. The training target is the relative completion
time of the fixed work. At runtime, Linux combines the resulting page scores
with capacity, migration cost, and any additional ranking inputs.

\section{Observation-Cost Methodology}
\label{app:prototype-overhead}

The measurement in
Figure~\ref{fig:microbenchmark-cost}\subref{fig:microbenchmark-overhead}
holds page placement fixed and disables migration. The same access stream
runs with STM disabled and with post-cache observation and the full
notification path enabled. Across eight equal-work windows within each run,
enabling STM increases CPU cycles by 1.2\%. Results are normalized to the
STM-disabled baseline; whiskers show one standard deviation across the
windows.

\section{Application Throughput}
\label{app:application-throughput}

\begin{table}[htbp]
  \centering
  \caption{Fit throughput gain over TPP in the online application runs.}
  \label{tab:application-throughput}
  \begin{tabular}{lrr}
    \toprule
    Workload & DRAM capacity & Throughput gain \\
    \midrule
    Redis & 25\% & 1.8\% \\
    Redis & 50\% & 6.0\% \\
    Redis & 75\% & 7.6\% \\
    \midrule
    llama.cpp & 15\% & 9.0\% \\
    llama.cpp & 25\% & 13.3\% \\
    llama.cpp & 50\% & 7.5\% \\
    \bottomrule
  \end{tabular}
\end{table}

Table~\ref{tab:application-throughput} reports throughput gains from the
same fixed-work runs as Figures~\ref{fig:redis-end-to-end}
and~\ref{fig:llama-end-to-end}, computed from the ratio of TPP's completion
time to Fit's. The runs include online controller and migration costs;
the llama.cpp workload measures the model's memory accesses.

\clearpage

\end{document}